# Selection of Compton-thick AGN from a hard photometric sample using *XMM–Newton* observations

Reham Mostafa,[1★] Matteo Guainazzi[2] and Alaa Ibrahim[3,4]

[1]*Physics Department, Science Faculty, Fayoum University, Fayoum 63514, Egypt*
[2]*European Space Research and Technology Centre, Neplerlaan 1 2201 AZ Noordwijk, The Netherlands*
[3]*Physics of Earth and Universe, Zewail City of Science and Technology, Giza 12578, Egypt*
[4]*Physics Department, Faculty of Science, Cairo University, Giza 12613, Egypt*

## ABSTRACT

We present a selection technique to detect Compton-thick (CT) active galactic nuclei (AGNs) in the 3XMM/SDSS-DR7 cross-correlation. A subsample of 3481 X-ray sources that are detected in the hard band (2–8 keV) and have photometric redshifts constitute our parent sample. We first applied an automated spectral-fitting procedure to select highly absorbed sources ($N_{\rm H} > 10^{23}$ cm$^{-2}$). We found 184 highly absorbed candidates. Then, we performed the Bayesian Monte Carlo Markov chains (MCMCs) selection technique to find CT AGNs. We also tested the MCMC selection technique by applying Monte Carlo simulations. We found that the method is accurate at 90 per cent independently of the nature of the underlying source. Our sample contains 52 *bona fide* CT AGNs. The CT AGNs were selected to have a range >0.75 of probability of being CT when either fitting with the two models Torus and MYTorus. About 75 per cent of CT AGNs in our sample had probabilities >90 per cent. From the spectral analysis, we significantly found an anticorrelation between the equivalent width of the neutral Fe K$\alpha$ line and the X-ray luminosity at 2–10 keV, the so-called X-ray Baldwin effect.

## 1 INTRODUCTION

Active galactic nuclei (AGNs) are energetic astrophysical sources powered by accretion on to supermassive black holes (SMBHs) at the centre of its host galaxy. They produce a gigantic amount of non-stellar emission that has been observed in the radio, microwave, infrared (IR), optical, ultraviolet (UV), X-ray, and gamma-ray wavebands. Various strategies are followed in different bands to identify and characterize these sources. In our work, we will discuss X-ray selected AGN. One of the main features that have been observed in the X-ray spectrum of AGN is the intrinsic power-law spectrum with photon index, Γ, ranging between 1.4 and 2.8, and its distribution in local AGN can be approximated by a Gaussian of mean 1.95 and standard deviation 0.15 (Nandra & Pounds 1994; Bianchi et al. 2009). Such a spectral shape is naturally explained by inverse Compton scattering of accretion disc photons by a hot corona of relativistic electrons (Titarchuck 1994). The intrinsic power law is often obscured by material between the accretion disc and the observer (Awaki et al. 1991; Bianchi et al. 2012). Systems where this occurs are called obscured AGN. The obscuring medium is composed of dust and/or gas. In this paper, we consider AGN to be obscured if the column density $N_{\rm H} \geq 10^{22}$ cm$^{-2}$. The impact of obscuration in the X-ray band is a function of rest-frame energy, with lower energy X-ray photons more easily absorbed than higher energy X-ray photons (Matt 2002). According to the unified model of AGN (Antonucci 1993; Netzer 2015), the accretion disc is surrounded by a geometrically and optically thick dusty and molecular torus (dusty torus). From the very centre to host-galaxy scales the main AGN structures are the accretion disc and the corona, the broad-line region (BLR), the torus, and the narrow-line region (NLR; see Ramos Almeida & Ricci 2017). It is likely that X-ray obscuration is produced by multiple absorbers on various spatial scales, mostly associated with the torus and the BLR (Bianchi et al. 2012; Ramos Almeida & Ricci 2017).

Most AGN are obscured in the X-ray band by large amounts of gas and dust (Hasinger 2008). If the X-ray obscuring matter has a column density $N_{\rm H} < 1.5 \times 10^{24}$ cm$^{-2}$ (the inverse of the Thomson cross-section for electrons, $\sigma_{\rm T}{}^{-1}$) the dominant X-ray absorption mechanism is photoelectric absorption, and hard X-rays (2–10 keV) can still be detected through the obscuring matter. However, for column densities above $10^{24}$ cm$^{-2}$ the attenuation of X-rays is dominated by Compton scattering on electrons rather than photoelectric absorption. These sources are called Compton-thick (CT) AGN. If the column density does not exceed a value of order $10^{25}$ cm$^{-2}$, then the nuclear radiation is still visible above 10 keV, and the source is called mildly CT. For higher column densities (heavily CT), the entire high-energy spectrum is downscattered by Compton recoil and hence depressed over the entire X-ray energy range (Comastri 2004).

CT matter can still be detected through a strong iron K$\alpha$ line, which is expected to be produced both by transmission through

★ E-mail: rma04@fayoum.edu.eg (RM); Matteo.Guainazzi@esa.int (MG)

(Leahy & Creighton 1993) and/or reflection by absorbing gas (Matt, Brandt, & Fabian 1996). Iron K$\alpha$ line observed in AGN spectra is composed by a broad component with full width at half-maximum (FWHM) $\geq$ 30 000 km s$^{-1}$ which is thought to be formed close to the black hole in the accretion disc (Fabian et al. 2000) or to be related to the partially covering warm absorbers in the line of sight (Miyakawa et al. 2012; Ricci et al. 2013). The other component is narrow, with FWHM $\sim$ 2000 km s$^{-1}$ (Shu, Yaqoob & Wang 2010) and peaks at 6.4 keV (Yaqoob & Padmanabhan 2004). This line is produced via fluorescence in cold neutral material that has been identified as circumnuclear matter located no closer than several thousand gravitational radii from the SMBH, most likely be the molecular torus (Nandra 2006). The line equivalent width (EW) of the iron line reaches values of order 1 keV for $N_{\rm H}$ $\sim$ 10$^{24}$ cm$^{-2}$ and larger values up to several keV can be obtained for high inclination angles and small torus opening angles (Levenson et al. 2002; Comastri 2004). Compton reflection produces also a continuum component, which is characterized by a broad bump peaking around 20−30 keV (George & Fabian 1991), while the 2–10 keV band is well approximated by a flat power law ($\Gamma \sim 0$).

An anticorrelation between the EW of the narrow component of the iron K$\alpha$ line and the X-ray luminosity was reported using *Ginga* data (Iwasawa & Taniguchi 1993). It is called Iwasawa–Taniguchi effect or X-ray Baldwin effect. The anticorrelation between the EW of a line and the luminosity in AGN was early detected in the UV by Baldwin (1977). The interpretation of Baldwin effect depends on the region in the nuclear environment where the line originates (Ricci et al. 2013).

Corral et al. (2014) presented an automated method for selecting highly absorbed candidate AGN using a sample of 1015 sources detected in the hard band (2–8 keV) from the XMM/SDSS cross-correlation presented in Georgakakis & Nandra (2011). They found 81 candidate highly obscured sources and after applying detailed manual spectral fits, 28 sources identified, exhibiting a column density of $N_{\rm H}$ > 10$^{23}$ cm$^{-2}$. Another efficient way for selecting CT AGN was applied by Akylas et al. (2016). Their selection technique depends on the use of Bayesian statistics to estimate the probability distribution of a source being CT. They use a sample of about 604 sources selected from the 70-month *Swift*-BAT all-sky survey in the 14–195 keV band and found 53 possible CT sources (probability range 3 per cent–100 per cent), representing ~7 per cent of the sample. A similar procedure described in Lanzuisi et al. (2018), which combines a physically motivated model for the X-ray emission with Monte Carlo Markov chain (MCMC) parameter estimation techniques and the use of the full probability distribution function (PDF) of the column density to select CT AGN. They selected 67 CT AGN candidates among the 1855 extragalactic sources from the *Chandra* COSMOS Legacy point source catalogue. In this paper, we will use these techniques to identify our sample of candidate heavily obscured and CT AGN.

The goal of our work is to search for the possible CT AGN present in a small sample of AGN taken from *XMM–Newton* serendipitous X-ray survey. We present a fully automated selection technique of highly obscured sources ($N_{\rm H}$ > 10$^{23}$ cm$^{-2}$), and then apply a Bayesian approach to derive the probability distribution of a source being a CT. In Section 2, we describe the sample selection and Bayesian approach technique. Section 3 applies Monte Carlo simulations to test the selection method. Section 4 presents the spectral analysis of physically motivated models of torus reflection (Murphy & Yaqoob 2009), and Section 5 presents the possible correlations in our CT sample.

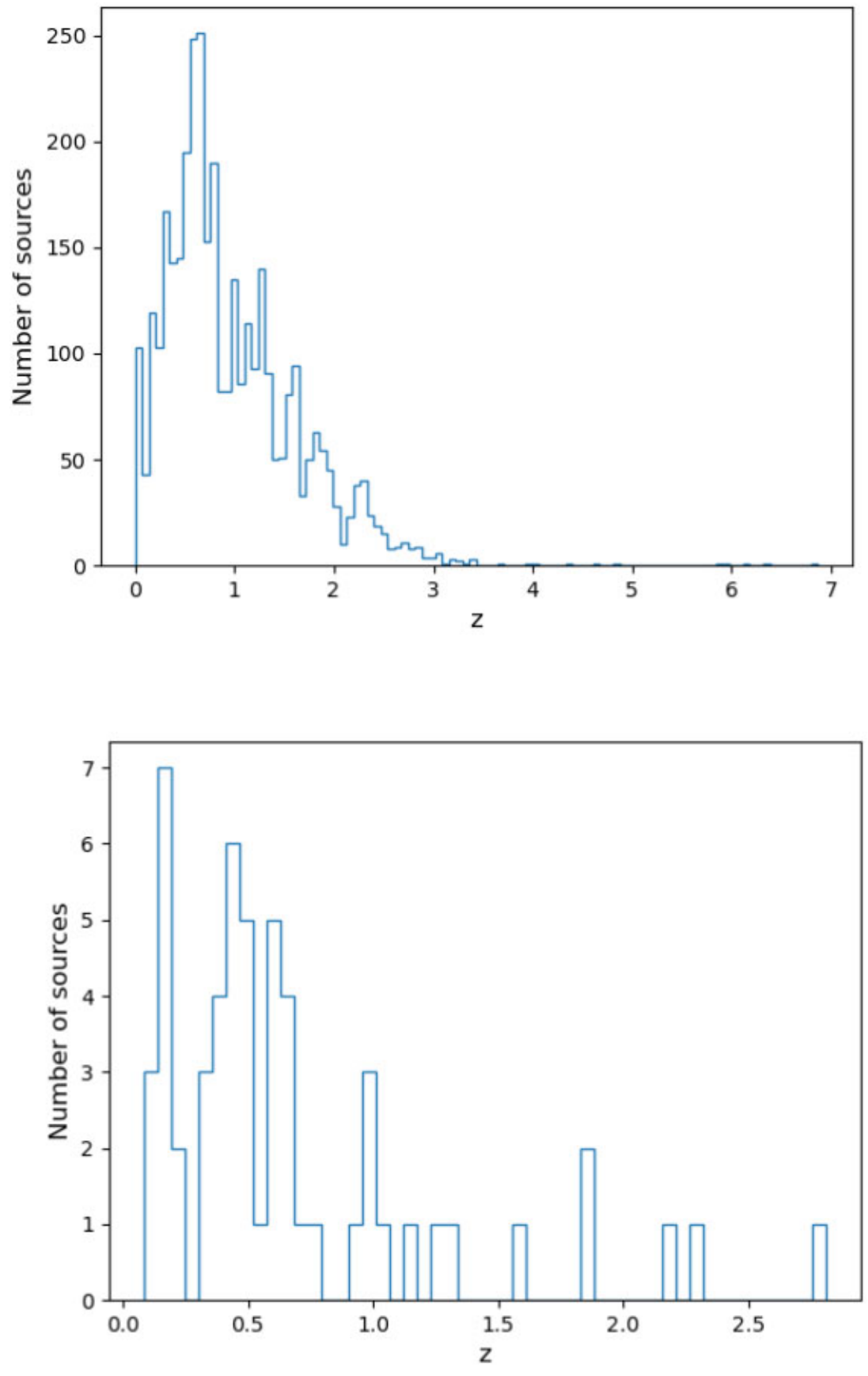

**Figure 1.** Redshift distribution of the parent sample (top) and the *bona fide* CT total sample (bottom).

## 2 SAMPLE SELECTION

We start with the X-ray catalogue XMM/SDSS-DR7 cross-correlation presented by Georgakakis & Nandra (2011). The XMM/SDSS is a serendipitous X-ray survey based on archival *XMM* observations that overlap with the SDSS DR7 footprint. The catalogue contains about 40 000 X-ray point sources which detected over a total area of 122 deg$^2$. We select the X-ray sources that are detected in the hard band (2–8 keV) and have photometric redshifts. This corresponds to 4288 sources. From this sample of hard X-ray sources, we further selected a sample of 3481 sources matched with the *XMM–Newton* spectral-fit Z catalogue (XMMFITCAT-Z; Ruiz, Georgantopoulos & Corral 2021). This catalogue contains spectral fitting results for 22 677 unique sources. It is produced by applying automated spectral fits to the EPIC spectral data contained within the *XMM–Newton* serendipitous source catalogue and using redshifts contained in the *XMM–Newton* Photo-Z catalogue (XMMPZCAT; Ruiz et al. 2018). 2740 of the 3481 sources have photometric redshifts and 741 sources have spectroscopic redshifts (Data details of the parent sample is presented as online supplementary material). The redshift distribution of the parent sample is shown in Fig. 1. A detailed description of the spectral-fit data base for this sample is presented on the project web page.[1]

The spectra of all sources in the sample are extracted by the automatic data reduction pipeline on which the production of the 3XMM-DR4 catalogue is based (Watson et al. 2009). We used precalculated redistribution matrices and effective area files provided by the *XMM–Newton* Science Operation Centre. We binned the spectra to 1 count/bin and used Cash statistics C-stat (Cash 1979) to perform the fitting. The spectra from the three EPIC instruments PN, MOS1, and MOS2 are fitted simultaneously with all the astrophysical parameters tied together among the three instruments. We also added a cross-normalization factor to constrain the difference in flux calibration for the different instruments. We fixed one of the three cross-normalization factors to one and left the other two factor values free to vary during the fit. Spectral analysis was carried out with XSPEC v12.9.1p (Arnaud 1996).

[1] https://xraygroup.astro.noa.gr/sites/prodex/xmmfitcatz_access.html

**Table 1.** Spectral models used in the analysis.

| Model | Used in section | XSPEC version | Energy (keV) |
|---|---|---|---|
| Pow | 2.1 | 12.9.1 | 2–10 |
| wabs∗zwabs∗(pow + zgaus) | 2.1 | 12.9.1 | 0.5–10 |
| wabs∗(zwabs∗(pow + zgaus)+pow) | 2.1 | 12.9.1 | 0.5–10 |
| wabs∗(zpow + torus + apec) | 2.2 | 12.9.1 | 0.5–10 |
| phabs∗(zpow∗MYtorusZ + constant∗MTtorusS + zgaus + zpow + apec) | 4 | 12.9.1 | 0.5–10 |

## 2.1 Selection of highly absorbed candidates

Following Corral et al. (2014), we perform an automated selection technique to select the highly absorbed candidates from our parent sample. This technique depends on four different automated selection criteria, which can be used to identify spectral properties of the most highly absorbed AGN.

Two of the criteria are based on spectral fitting performed by us. Two other criteria employ spectral-fitting results produced from the XMMFITCAT-Z catalogue. It uses four models, two simple (absorbed power law and absorbed thermal) models for sources with spectra corresponding to a single instrument, more than 50 net counts in the 0.5–10 keV band, and two additional complex (absorbed double power law and absorbed thermal plus power law) models for sources with total EPIC counts higher than 500 net counts in the same energy band. The spectral models used in the automated spectral-fitting include the effects of redshift and Galactic absorption. A list of the spectral models that we have employed in this work are shown in Table 1.

According to the Corral et al. (2014) selection technique, the source is considered highly absorbed candidate if it fulfills any of the following criteria:

1) FLATH sample, sources with photon index $< 1.4$ in the 2.0–10.0 keV rest-frame energy band fitted with simple power law. First, we apply an automated procedure to fit all sources using a simple power-law model. The fitting is performed in the hard (2–10 keV) band. We exclude the spectra with fewer than 20 counts in the hard band since we cannot compute the value of photon index. Then, we select sources with $\Gamma < 1.4$ at the 90 per cent confidence level. 54 sources are selected as highly absorbed candidates.

2) FLATA sample, sources with photon index $< 1.4$ at the 90 per cent confidence level fitted with absorbed power law in the full band (0.5–10 keV). These photon index values are derived from the automated spectral-fitting results within 3XMM spectral-fit data base (XMMFITCAT-Z). 26 sources are selected as highly absorbed candidates.

3) HABS sample, sources with intrinsic $N_{\rm H} > 5 \times 10^{23}$ cm$^{-2}$ when fitted with absorbed power law in the full band (0.5–10 keV). The $N_{\rm H}$ values are available from the automated spectral-fitting results within 3XMM spectral-fit data base (XMMFITCAT-Z). Five sources are selected as highly absorbed candidates.

4) HEW sample, sources with EW $>$ 500 eV at the 90 per cent confidence level in the best-fitting model. This sample selection is slightly different from that used by Corral et al. (2014) in which they used the four models (absorbed power law, absorbed double power law, absorbed thermal plus power law, and absorbed blackbody plus power law) plus a narrow Gaussian emission line. For sources with more 500 counts, we fit the 0.5–10 keV spectra with a partially absorbed double power law plus a Gaussian line model. The Gaussian line included in those models to represent the Fe K$\alpha$ emission line had a fixed line rest-frame energy of 6.4 keV and width of 0.1 keV. Then, we select sources with either EW $>$ 500 eV at the 90 per cent confidence level or $N_{\rm H} > 10^{24}$ cm$^{-2}$ at the 90 per cent confidence level in the best-fitting model. We consider the iron line to be significant if $\Delta$C-stat $\geq$ 4.56. 102 sources are selected as highly absorbed candidates.

The final sample of highly absorbed candidates has 184 sources.

## 2.2 Bayesian approach

To avoid selecting the CT AGN depending on the value of $N_{\rm H}$ derived from phenomenological fits, we apply the MCMC method described by Akylas et al. (2016). We refitted all the 184 highly absorbed candidates extracted from the procedure described in section 2.1 with the Torus model described in Brightman & Nandra (2011) since it is simple and proper model for fitting spectra of heavily obscured sources. This model assumes a spherical torus with a biconical void and the line-of-sight $N_{\rm H}$ through the torus is constant (not depend on the inclination angle). The model also has a variable opening angle (26°–4°) and $N_{\rm H}$ reaches up to $10^{26}$ cm$^{-2}$. During the fit, we fixed the opening angle to 60° and the viewing angle to 80°.To this model, we add the `wabs` component to account for the absorption along the line of sight due to our Galaxy and an unabsorbed power-law component in order to describe the component that comes from scattered emission, if significantly required by the fit. An additional thermal emission component `apec` (Smith et al. 2001) is added if it is required by the fit in a significant level to account for the soft X-ray emission from the central region of the host galaxy. After performing the spectral fit with the Torus model in the (0.5–10 keV) band using the cash statistic and getting the best fit, we run an MCMC within XSPEC using the Goodman–Weare algorithm (Godman & Waare 2010) with $10^4$ steps to efficiently estimate the parameter space. Then, the $N_{\rm H}$ parameter for each candidate is marginalized to derive the PDF. From the probability distribution for each candidate, we can assign the probability of being CT AGN ($P_{\rm CT}$).

As a result, we found 69 sources that have a non-zero probability of being CT [$P_{\rm CT} > 0$] from all the highly absorbed candidates. We call this sample the *probable* CT sample. Since CT sources are reflection-dominated sources in the EPIC sensitive bandpass, the observed 2–10 keV luminosity is dominated by the torus component which can only provide a measurement of the reflected, and not of the nuclear, luminosity. In this case, a correction factor of 70 was applied to the observed 2–10 keV luminosity to infer the intrinsic luminosity of CT sources following Marinucci et al. (2012). In Table 1, we present the fitting results for the *probable* CT sample sources. The probability of being CT ($P_{CT}$) of CT candidates in our sample is shown in Fig. 2.

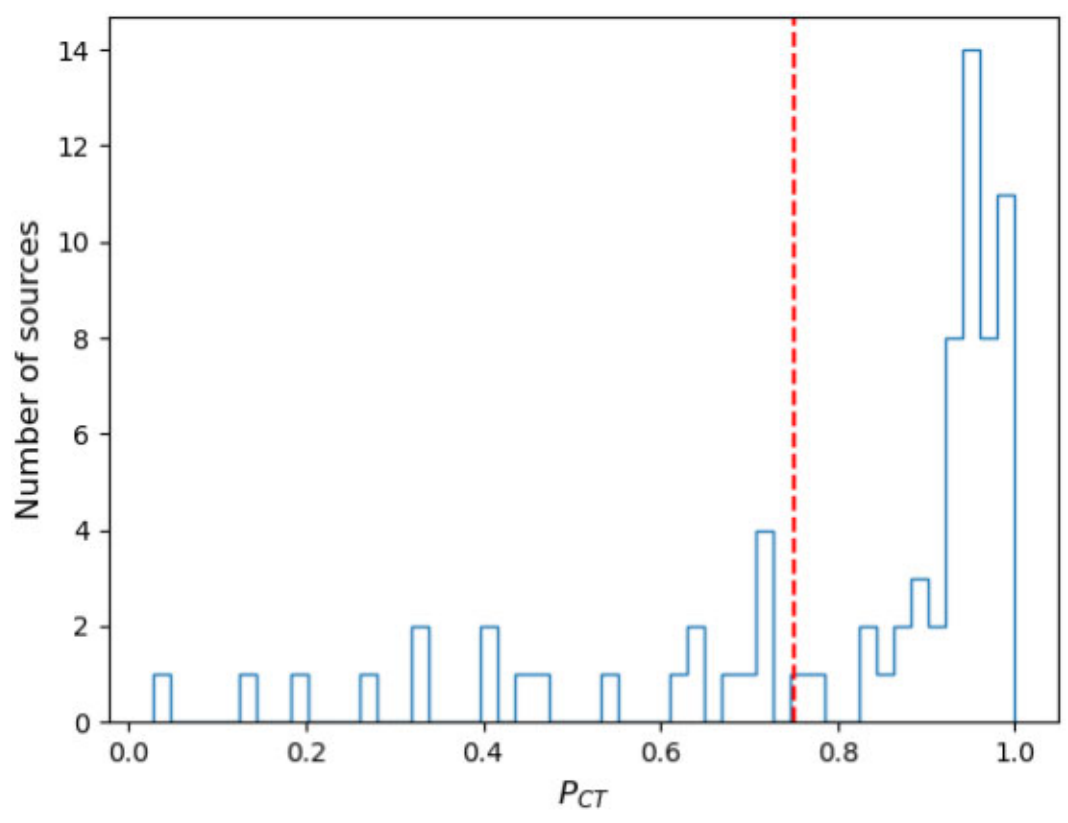

**Figure 2.** The probability of being CT distributions of the *probable* CT sample. The dashed line shows the threshold for selecting the *bona fide* CT sample.

There are three sources with significant `apec` component and temperature $\leq$1 keV at the 90 per cent confidence level (3XMMJ080622.9+152731, 3XMMJ115903.3+440149, and 3XMMJ224304.5−094241). For the case of 3XMMJ080622.9+152731, the `apec` component luminosity exceeds $10^{42}$ erg s$^{-1}$ but adding this component greatly improve the fit from C/dof = 5413.22/1072 to 1023.05/1070 so we retain this component, where dof is the number of degrees of freedom. We did not achieve a statistical improvement in the fit when adding an unabsorbed power law for the two sources 3XMMJ102408.3+040931 and 3XMMJ115903.3+440149.

## 3 MONTE CARLO SIMULATIONS

We need to test the goodness of our method of selection for CT AGN. A good way to follow is through Monte Carlo simulations. First, we define the best-fitting Torus model to base the simulation. Then, we create the simulated spectra using the command *fakeit* in XSPEC, taking into account the exposure time for each source, for all candidates in the highly absorbed sample (184 sources). Once created, the simulated spectra are analysed in the same way as real data. This allows us to estimate the probability of being CT and therefore how accurate our classification method is. Now, we can answer the two questions: how often is a CT object correctly identified as a CT source? And how often is a non-CT object erroneously identified as a CT source? We found that from 69 sources in the *probable* CT sample, 62 sources are identified as CT object by the spectral analysis of simulated data. Out of 97 non-CT highly absorbed sources in the parent sample, 5 sources are detected as CT object by the spectral analysis of simulated data. This means that the selection method is accurate at the 90 per cent level, with a possible level of residual contamination <10 per cent. We also found that the fraction of correctly identified CT sources is not dependent on the flux, luminosity, or net count rate in the 2–10 keV band.

An even more stringent criterion on $P_{CT}$ is applied to be sure that the sample is not contaminated with objects that are not CT AGN. We apply a conservative threshold of $P_{CT} \geq 0.75$ on the sample of 62 sources that were identified as CT by the spectral analysis of simulated data. This conservative threshold excludes 15 sources from the sample. Therefore, the *bona fide* CT Torus-selected sample is composed of 47 sources and are seen in bold in Table 2.

We plot the folded spectra and their $\Gamma$ and $N_H$ PDF for the 47 CT sources (see Fig. A1). The distribution of their photon index, inferred 2–10 keV luminosity, redshift, and column density is plotted in Fig. 3. For plotting those column density and photon index distributions, we exclude the lower limits.

## 4 MODELLING WITH MYTORUS MODEL

To study the properties of the *bona fide* CT sources, we use the model MYTorus (Murphy & Yaqoob 2009) since it takes into account Compton scattering properly. In order to avoid any inconsistencies in describing the properties of the torus as a result of using two different models for the torus reflection, we fit the X-ray spectra of all sources that are identified as CT object by the spectral analysis of simulated data in the *probable* CT sample (62 sources) with the MYTorus model and compute the $P_{CT}$ for each source by following the Bayesian approach described in Section 2.2. We could not use MYTorus model instead of Torus model when applying the selection technique MCMC described in Section 2.2 due to computational issues and to its complexity, overfitting some of the spectra in the original sample.

This model is suitable for modelling X-ray spectra from a toroidal reprocessor with a half-opening angle of 60° equivalent to a covering factor of $\Delta\Omega/4\pi = 0.5$. It covers the Compton-thin and CT regimes with column densities in the range $10^{22}$–$10^{25}$ cm$^{-2}$. We use Anders & Grevesse (1989) elemental cosmic abundances and photoelectric cross-sections from Verner & Yakovlev (1995) and Verner et al. (1996). The inclination angle between the observer's line of sight and the symmetry axis of the torus is given by $\theta_{obs}$, where $\theta_{obs} = 0°$ corresponds to a face-on observing angle and $\theta_{obs} = 90°$ corresponds to an edge-on observing angle.

Incident continuum photons, which do not intercept the torus, represent the zeroth order continuum component (direct or transmitted continuum) in MYTorus. It does not depend on the geometry or covering factor of the material out of the line of sight but it is a fraction of the input spectrum at a given energy and that fraction depends only on the optical depth of the torus gas at that energy. Photons that have been scattered in the medium at least once constitute the scattered continuum (reflection spectrum). If photons are absorbed at energies higher than the K-edge threshold energy of an atom or ion, fluorescent lines are emitted. In XSPEC, MYTorus is applied through the three model components: zeroth order continuum, Compton-scattered continuum, and the fluorescent line emission.

Besides the MYTorus model components, we include a Galactic absorption component (phabs in XSPEC), an unabsorbed power-law component (`zpowerlw` in XSPEC), a thermal component (apec in XSPEC), and two additional Gaussian lines to account for the ionized (He-like and H-like) iron emission lines. The latter three components are added only if the fit is significantly improved. The rest-frame energy of the ionized lines from He-like and H-like ions are fixed to 6.7 and 6.96 KeV, respectively, and their intrinsic width was fixed to 0.

In XSPEC the model is given by:

constant*phabs(zpowerlw*etable{mytorus_Ezero_v00.fits}+
constant*atable{mytorus_scatteredH500_v00.fits}
+zpowerlw+zgaus+zgaus+apec)

The normalization parameter of the scattered continuum component is tied to the normalization of the intrinsic power-law continuum ($A_i = A_s$). Also, the photon index, the line-of-sight column density

**Table 2.** MCMC spectral-fitting results of the ***probable*** CT sample.

| Source (1) | $P_{CT}$ ($P_{CT(fa)}$) (2) | $N_H$ ($10^{22}$) cm$^{-2}$ (3) | $\Gamma$ (4) | $\log_{10} L_{2-10}$ keV (5) | Flux ($10^{-14}$) erg cm$^{-2}$ s$^{-1}$ (6) | Z (7) | Sample (8) | Target (9) |
|---|---|---|---|---|---|---|---|---|
| **3XMMJ102408.3+040931** | **0.88(0.85)** | $100.1^{+37.8}_{-29.0}$ | **1.00** | **45.42** | **1.99** | **2.3 ± 1.64** | **3** | **Zw3146** |
| 3XMMJ111715.8+175741 | 0.13(0.67) | $46.4^{+24.7}_{-17.9}$ | 1.33 ± 0.11 | 46.26 | 29.34 | 0.60 ± 0.18 | 1, 2 | DP_Leo(NGC 3608) |
| 3XMMJ115903.3+440149 | 0.68(0.51) | $1.5^{+1.2}_{-0.7}$ | 1.00 | 44.26 | 2.19 | 0.22 | 2 | NGC 4013 |
| 3XMMJ121010.7+392300 | 0.19(0.41) | $28.3^{+13.2}_{-5.7}$ | $2.00^{+0.57}_{-0.48}$ | 44.04 | 2.26 | 0.18 ± 0.08 | 2 | NGC 4151 |
| **3XMMJ121051.2+393206** | **0.76(0.42)** | $102.0^{+24.7}_{-26.0}$ | **1.95 ± 0.26** | **46.41** | **6.87** | **1.57 ± 0.08** | **1,2** | **NGC 4151** |
| 3XMMJ122558.8+332823 | 0.33(0.26) | $30.3^{+13.7}_{-7.7}$ | $2.14^{+0.50}_{-0.40}$ | 44.38 | 1.42 | 0.32 ± 0.07 | 2 | NGC 4395 |
| 3XMMJ151616.1+000148 | 0.54(0.84) | $19.3^{+289.7}_{-8.5}$ | 1.00 | 45.53 | 5.14 | 0.65 ± 0.10 | 2 | RXCJ1516.3 + 0005 |
| **3XMMJ080622.9+152731** | **0.99(0.99)** | $158.0^{+50.1}_{-10.6}$ | **>2.46** | **47.03** | **4.14** | **2.19 ± 1.39** | **1** | **RX J0806.3 + 1527** |
| **3XMMJ085707.9+275319** | **0.78(0.70)** | $51.3^{+41.9}_{-24.5}$ | **2.37 ± 0.13** | **45.20** | **6.57** | **0.34 ± 0.31** | **1** | **3C 210** |
| **3XMMJ100056.5+554059** | **0.92(0.93)** | $112.5^{+139.0}_{-47.5}$ | $1.78^{+0.12}_{-0.06}$ | **46.64** | **18.06** | **1.04** | **1** | **SN2001ci(NGC 3079)** |
| 3XMMJ100114.8+020208 | 0.63(0.54) | $56.2^{+25.5}_{-18.2}$ | $2.35^{+0.18}_{-0.15}$ | 46.12 | 5.54 | 0.98 ± 0.20 | 1 | CID-42 |
| **3XMMJ100159.7+022641** | **0.94(0.95)** | $158.1^{+69.1}_{-46.2}$ | $2.06^{+0.13}_{-0.10}$ | **46.86** | **7.00** | **1.88 ± 0.13** | **1** | **COSMOS_FIELD** |
| 3XMMJ105106.3+573435 | 0.41(0.24) | $14.0^{+5.6}_{-6.0}$ | >1.29 | 45.12 | 3.20 | 0.45 ± 0.09 | 1 | Lockman_Hole |
| **3XMMJ121047.2+392313** | **0.96(0.94)** | **>240.2** | $1.87^{+0.13}_{-0.11}$ | **46.06** | **4.25** | **1.13 ± 0.22** | **1** | **NGC 4151** |
| 3XMMJ140921.7+261317 | 0.03(0.93) | $3.0^{+2.2}_{-1.1}$ | 1.00 | 45.73 | 6.89 | 0.68 ± 0.13 | 1 | PG1407 + 265 |
| **3XMMJ143031.3−000907** | **0.95(0.91)** | $172.9^{+98.6}_{-50.3}$ | $2.17^{+0.21}_{-0.19}$ | **47.25** | **8.38** | **2.81** | **1** | **SDSSJ1430−0011** |
| 3XMMJ030705.8−000009 | 0.33(0.49) | $30.7^{+22.8}_{-10.4}$ | 2.73 ± 0.10 | 44.81 | 4.37 | 0.27 | 4 | S2F1a |
| **3XMMJ083734.6+060644** | **1(0.92)** | **>168.2** | **2.12 ± 0.13** | **45.70** | **3.04** | **0.77** | **4** | **PSRB0834+06** |
| **3XMMJ090044.9+385521** | **0.83(0.86)** | $39.6^{+107.9}_{-21.7}$ | **2.21 ± 0.12** | **44.58** | **10.39** | **0.14** | **4** | **IRAS 08572+3915** |
| 3XMMJ092040.4+302520 | 0.27(0.0) | $7.2^{+6.0}_{-2.3}$ | 2.05 ± 0.16 | 44.96 | 4.11 | 0.32 ± 0.09 | 4 | A781 |
| 3XMMJ092318.5+511400 | 0.94(0.0) | 112.2 | 2.08 ± 0.10 | 44.65 | 6.11 | 0.20 ± 0.43 | 4 | 1E 0919+515 |
| 3XMMJ093535.4+611919 | 0.71(0.71) | $38.1^{+36.5}_{-19.8}$ | 2.40 ± 0.08 | 44.69 | 6.05 | 0.20 | 4 | UGC05101 |
| **3XMMJ095942.9+251524** | **0.93(0.95)** | **>40.9** | $2.37^{+0.14}_{-0.10}$ | **44.55** | **8.09** | **0.16 ± 0.08** | **4** | **cfrs10h** |
| **3XMMJ100101.5+250904** | **1(0.84)** | **>3117.4** | **2.18 ± 0.11** | **45.34** | **3.55** | **0.50** | **4** | **cfrs10h** |
| **3XMMJ100336.5+325154** | **1(0.85)** | **>23.1** | **2.02 ± 0.08** | **44.13** | **6.31** | **0.11** | **4** | **RXJ1003.0 + 3254** |
| **3XMMJ102224.2+383504** | **0.93(0.92)** | **>51.91** | **1.98 ± 0.08** | **44.24** | **15.68** | **0.08 ± 0.26** | **4** | **RXCJ1022.0 + 3830** |
| **3XMMJ102836.0+240402** | **0.95(0.93)** | $145.9^{+80.5}_{-67.3}$ | **2.21 ± 0.11** | **45.99** | **15.29** | **0.59 ± 0.08** | **4** | **87GB1025 + 2416** |
| **3XMMJ103446.8+575128** | **0.97(0.97)** | **>1808.0** | **2.20 ± 0.12** | **45.81** | **5.07** | **0.68 ± 0.87** | **4** | **LH-NW2** |
| 3XMMJ103523.8+392752 | 0.46(0.0) | $2.9^{+0.9}_{-0.8}$ | >2.62 | 45.16 | 4.59 | 0.33 | 4 | REJ1034 + 396 |
| **3XMMJ104048.5+061819** | **0.94(0.96)** | $158.0^{+100.5}_{-62.1}$ | **1.90 ± 0.05** | **44.82** | **14.26** | **0.16** | **4** | **4C06.41** |
| **3XMMJ104421.8+063545** | **0.99(0.94)** | **>218.5** | **2.40 ± 0.14** | **45.53** | **3.34** | **0.60 ± 0.90** | **4** | **NGC 3362** |
| **3XMMJ111238.1+132245** | **0.96(0.97)** | **>2816.1** | **1.45 ± 0.09** | **45.54** | **9.53** | **0.43** | **4** | **Abell 1201** |
| **3XMMJ111247.5+132419** | **0.94(0.93)** | **112.0** | **1.66 ± 0.08** | **44.09** | **3.55** | **0.14 ± 0.28** | **4** | **Abell 1201** |
| **3XMMJ112253.5+052025** | **0.95(0.94)** | $39.8^{+65.5}_{-23.9}$ | **2.04 ± 0.10** | **44.48** | **11.96** | **0.12 ± 0.08** | **4** | **3C 257** |
| **3XMMJ113322.8+663326** | **0.96(0.94)** | $176.3^{+119.8}_{-87.3}$ | **2.19 ± 0.10** | **46.38** | **23.65** | **0.71 ± 0.70** | **4** | **A1302** |
| 3XMMJ115339.3+224920 | 0.71(0.77) | $46.9^{+49.6}_{-20.1}$ | 2.45 ± 0.11 | 46.09 | 6.82 | 0.77 ± 0.08 | 4 | A1413offset |
| **3XMMJ120049.8+341852** | **0.96(0.91)** | **>117.5** | **2.18 ± 0.12** | **45.27** | **2.45** | **0.56 ± 0.63** | **4** | **6C1200 + 3416** |
| **3XMMJ121844.2+055714** | **0.96(0.95)** | $112.3^{+111.7}_{-57.8}$ | **1.86 ± 0.10** | **44.61** | **6.03** | **0.20 ± 0.10** | **4** | **NGC 4261** |
| **3XMMJ121955.1+054643** | **0.97(0.96)** | **>163.2** | **2.06 ± 0.08** | **44.97** | **3.06** | **3.07 ± 0.12** | **4** | **NGC 4261** |
| **3XMMJ125331.9+100532** | **0.88(0.82)** | $66.9^{+73.9}_{-35.3}$ | **2.12 ± 0.09** | **45.68** | **11.98** | **0.43 ± 0.13** | **4** | **IRAS F12514+1027** |
| **3XMMJ122935.6+075326** | **0.96(0.33)** | **158.9** | **2.02 ± 0.11** | **45.20** | **1.96** | **0.58 ± 0.13** | **4** | **NGC 4472** |
| **3XMMJ123047.0+105606** | **0.93(0.77)** | $112.1^{+167.6}_{-68.5}$ | **2.56 ± 0.09** | **45.45** | **2.30** | **0.66 ± 0.22** | **4** | **LBQS1228 + 1116** |
| **3XMMJ123121.4+135955** | **1(0.92)** | **157.9** | **1.97 ± 0.11** | **46.00** | **12.63** | **0.60** | **4** | **Virgo5** |
| **3XMMJ123140.5+254342** | **0.96(0.95)** | **>63.1** | **1.72 ± 0.10** | **45.33** | **5.09** | **0.45** | **4** | **NGC 4494** |
| 3XMMJ123601.9+263509 | 0.72(0.91) | $20.0^{+16.7}_{-12.1}$ | $2.07^{+0.15}_{-0.12}$ | 45.60 | 4.90 | 0.57 ± 0.23 | 4 | NGC 4555 |
| 3XMMJ130803.3+292909 | 0.94(0.0) | $112.0^{+167.9}_{-72.6}$ | 1.90 ± 0.07 | 45.03 | 7.55 | 0.28 ± 0.08 | 4 | SA57 |
| 3XMMJ131240.8+230924 | 0.72(0.97) | $27.7^{+26.4}_{-12.7}$ | 2.36 ± 0.07 | 45.34 | 8.39 | 0.34 | 4 | FBQSJ131213.5 + 23195 |
| **3XMMJ132529.9+320117** | **0.92(0.28)** | $112.2^{+112.9}_{-60.9}$ | **1.68 ± 0.09** | **46.06** | **25.73** | **0.48** | **4** | **4C + 32.44** |
| **3XMMJ132930.3+114857** | **0.93(0.96)** | $126.2^{+89.7}_{-54.7}$ | **2.23 ± 0.12** | **45.48** | **6.08** | **0.51** | **4** | **NGC 5171Group** |
| **3XMMJ133654.6+514611** | **0.95(0.98)** | $157.4^{+87.0}_{-73.7}$ | **2.14 ± 0.09** | **45.76** | **6.86** | **0.63 ± 0.10** | **4** | **UXUma** |
| **3XMMJ140638.2+222117** | **0.98(0.98)** | **>1136.4** | **1.86 ± 0.08** | **45.65** | **16.47** | **0.36** | **4** | **PG1404 + 226** |

**Table 2** – *continued*

| Source (1) | $P_{CT}$ ($P_{CT(fa)}$) (2) | $N_H$ ($10^{22}$) cm$^{-2}$ (3) | Γ (4) | $\log_{10} L_{2-10}$ keV (5) | Flux ($10^{-14}$) erg cm$^{-2}$ s$^{-1}$ (6) | Z (7) | Sample (8) | Target (9) |
|---|---|---|---|---|---|---|---|---|
| **3XMMJ140726.5+524710** | **0.90(0.74)** | $102.3^{+172.5}_{-74.2}$ | **2.18 ± 0.13** | **46.37** | **10.11** | **0.96** | **4** | **RXSJ140654.5 + 525 316** |
| **3XMMJ140858.3+261 239** | **0.86(0.94)** | $79.4^{+49.3}_{-29.3}$ | **2.32 ± 0.06** | **45.55** | **7.73** | **0.45** | **4** | **PG1407 + 265** |
| 3XMMJ123038.4+214056 | 0.24(0.0) | >2.75 | $3.38^{+0.81}_{-1.00}$ | 45.12 | 7.47 | 0.27 | 4 | NGPRift |
| 3XMMJ142259.8+193536 | 0.93(0.0) | $125.2^{+163.7}_{-71.7}$ | 1.95 ± 0.08 | 45.20 | 13.94 | 0.25 ± 0.14 | 4 | 3C300 |
| **3XMMJ144250.0−004249** | **0.94(0.93)** | **>478.3** | $2.14^{+0.15}_{-0.13}$ | **46.43** | **5.17** | **1.31 ± 0.10** | **4** | **RXJ~1442−0039** |
| 3XMMJ143013.8−001020 | 0.71(0.82) | >4.5 | 2.08 ± 0.10 | 45.14 | 32.52 | 0.15 ± 0.08 | 4 | SDSSJ1430−0011 |
| **3XMMJ151640.1−010212** | **0.88(0.91)** | $112.0^{+181.3}_{-75.5}$ | **1.65 ± 0.07** | **44.69** | **12.47** | **0.15 ± 0.08** | **4** | **RXC J1516.5−0056** |
| **3XMMJ153248.5+043948** | **0.96(0.99)** | **>237.6** | **2.24 ± 0.12** | **46.14** | **4.99** | **0.97** | **4** | **MKW9** |
| 3XMMJ160731.5+081202 | 0.63(0.0) | $17.1^{+49.2}_{-13.1}$ | 2.60 ± 0.13 | 44.70 | 5.94 | 0.20 ± 0.09 | 4 | XBSJ160645.9+081525 |
| **3XMMJ224304.5−094241** | **1(0.95)** | **>144.8** | **2.15 ± 0.08** | **44.90** | **2.63** | **0.38 ± 0.30** | **4** | **MACSJ2243.3−0935** |
| **3XMMJ234715.9+005602** | **0.91(0.97)** | $83.9^{+183.2}_{-38.5}$ | **2.35 ± 0.09** | **45.61** | **7.81** | **0.48 ± 0.12** | **4** | **1AXGJ234725+0053** |
| **3XMMJ011829.6+004548** | **0.84(0.81)** | $78.9^{+224.6}_{-45.7}$ | **2.22 ± 0.07** | **45.15** | **23.45** | **0.18 ± 0.11** | **4** | **011 905.14 + 003745.0** |
| **3XMMJ083827.1+255050** | **0.93(0.95)** | $305.6^{+245.7}_{-112.5}$ | **1.92 ± 0.13** | **46.64** | **25.48** | **0.94 ± 0.09** | **4** | **NGC 2623** |
| 3XMMJ085846.5+274534 | 0.45(0.78) | $17.2^{+17.4}_{-7.5}$ | 2.35 ± 0.09 | 44.95 | 4.35 | 0.30 ± 0.11 | 4 | 3C 210 |
| **3XMMJ115010.2+550709** | **0.95(0.93)** | **>339.7** | **1.92 ± 0.13** | **47.02** | **9.66** | **1.83** | **4** | **NGC 3921** |
| **3XMMJ130307.7−022427** | **0.89(0.57)** | $90.7^{+173.4}_{-59.2}$ | $2.08^{+0.17}_{-0.14}$ | **46.60** | **8.94** | **1.27 ± 0.12** | **4** | **RXC J1302.8−0230** |
| **3XMMJ130336.8+673027** | **0.95(0.97)** | **>163.4** | **2.25 ± 0.05** | **45.25** | **17.30** | **0.23 ± 0.10** | **4** | **A1674** |
| **3XMMJ134859.4+601457** | **1(0.75)** | **>109.3** | **2.13 ± 0.10** | **45.40** | **6.57** | **0.41** | **4** | **NGC 5322** |

**Notes.** *Columns:* (1) Source name in the 3XMM-DR4 catalogue. (2) Probability of being CT from Torus model (fake probability of being CT from Monte Carlo simulations). (3) The best-fitting column density. (4) The best-fitting photon index. (5) The logarithm of the inferred 2–10 keV luminosity. (6) 2–10 keV absorbed flux. (7) Photometric redshift and redshift values without error represent the spectroscopic redshift. (8) Detected sample for highly absorbed candidates. (9) Target. All errors are at 1$\sigma$ confidence level. Sources in bold are the *bona fide* CT Torus-selected sample.

$N_{H,Z}$, redshift z, and inclination angle ($\theta_{obs}$ fixed to 80°) of the MYTorus components are tied together. The relative normalization of the scattered continuum $A_s$ is modelled by a constant model component.

Since the emission line component in MYTorus model includes all the lines produced in the optically thick reprocessor, among them the Fe K$\alpha$ fluorescent line whose properties we want to explore, we cannot extract the observed properties of the line directly using this model. Moreover, MYTorus is not able to account in a self-consistent way with the ionization status of the gas. If the gas is slightly ionized, the centroid of the fluorescent line will be shifted to higher energy with respect to the cold case assumed by MYTorus. Therefore, in order to derive observables related to the iron lines, we performed a new set of spectral fit after replacing the emission line component of MYTorus, in which the velocity broadening is implemented with the XSPEC Gaussian convolution model `gsmooth`, with a simple Gaussian profile. The centroid energy was left free in the fit, and the width fixed to the best-fitting value in the gsmooth component of the baseline model. We fix the centroid energy to 6.4 keV if its value is lower than 6.4 keV at the 1$\sigma$ level.

The spectral-fitting results corresponding to the 62 *probable* CT sources are shown in Table 3. For one source, where the photon index pegs at the minimum allowed value (1.4), we fix the photon index to $\Gamma$ = 1.9 to better constrain the intrinsic absorption. Adding an apec model component significantly improves the fit in the three sources 3XMMJ080622.9+152731, 3XMMJ100101.5+250904, and 3XMMJ224304.5−094241 with temperature ≤1 keV at the 90 per cent confidence level. As in the case of fitting the spectrum of 3XMMJ080622.9+152731 with torus model the luminosity of apec component exceeds $10^{42}$ erg s$^{-1}$ but adding this component greatly improves the fit from C/dof = 6516.82/1071 to 1018.96/1069 so we retain the component. The addition of the unabsorbed power-law component significantly improves the fit except for the source 3XMMJ102408.3+040931. There are 21 sources and 2 sources for which the normalization of the scattering component is not constrained, marked with a ‘s’ in Table 3, and is zero, marked with a ‘z’ in Table 3, respectively. Of the 62 *probable* CT sources, 5 sources are found to have a significant neutral Fe K$\alpha$ emission line with EW up to 0.6 keV. The ionized iron emission line at 6.96 KeV is detected in one source 3XMMJ133654.6+514611 with EW 0.33 ± 0.001 keV. Then, we applied the MCMC method to the 62 *probable* CT sources as described in Section 2.2. We ended up with 39 *bona fide* CT sources, with a threshold $P_{CT} \geq 0.75$. This *bona fide* CT MYTorus-selected sample is indicated in bold in Table 3. The $P_{CT}$ distribution for both *bona fide* CT Torus-selected and *bona fide* CT MYTorus-selected samples is presented in Fig. 4.

Torus and MYTorus correspond to different geometries. It is therefore possible that an intrinsically CT AGN will be recognized as such only when the corresponding X-ray spectrum is fit with the model that best describes the geometry. So, we consider an object as *bona fide* CT when either of the two models yields a $P_{CT}$ higher than 0.75. According to this criterion, we got a total of 52 *bona fide* CT and hereafter we called this sample as the *bona fide* CT total sample.

The distributions of the photon index and hard X-ray luminosities of sources in the *bona fide* CT total sample are presented in Figs 5 and 6. Before fitting the distribution with Gaussian, we exclude the sources with lower limits and with $\Gamma = 1$ and $\Gamma = 3$ in the case of fitting with Torus model and the sources with $\Gamma = 2.6$ and $\Gamma$ fixed to 1.9 in the case of fitting with the MYTorus model. The mean value of $\Gamma$ for the *bona fide* CT total sample is 2.07 ± 0.11 with a standard deviation of $\sigma = 0.24$ and 2.06 ± 0.12 with a standard deviation of $\sigma = 0.21$ when fitting with Torus model and MYTorus model, respectively.

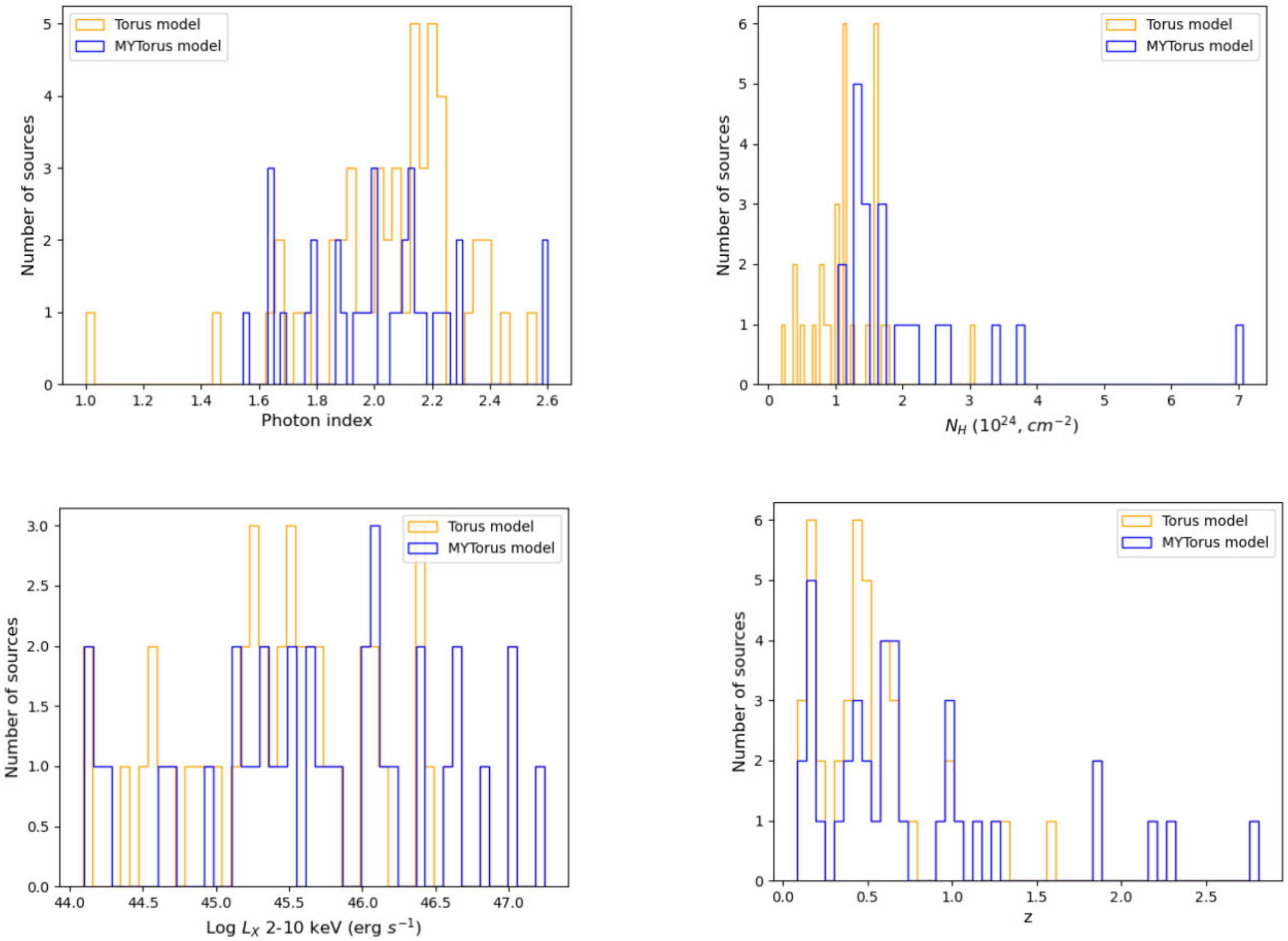

**Figure 3.** The distributions for the photon index, inferred 2–10 keV luminosity, column density, and redshift, respectively, for the *bona fide* CT Torus-selected sample and the *bona fide* CT MYTorus-selected sample.

### 4.1 Possible correlations in our *bona fide* CT samples

In this section, we check if there is any correlation between the X-ray spectral properties of CT AGN (column density, photon index, and luminosity) for the *bona fide* CT Torus-selected, *bona fide* CT MYTorus-selected, and *bona fide* CT total samples. For the *bona fide* CT total sample, we use the parameters taken from the Torus model to check for the possible correlations hereafter. A summary of those correlations is listed in Table 4. In the presence of censored data, we used the Akritas–Thiel–Sen (ATS) Kendall $\tau$-rank correlation test (Akritas, Murphy & LaValley 1995; Helsel 2005; Feigelson & Babu 2012). We performed ATS Kendall $\tau$-rank correlation test using r and the NADA[2] package. In addition, we used the Bayesian fitting method developed in Kelly 2007 and implemented in LINMIX-ERR package[3] which incorporates errors in both independent and dependent variables as well as non-detections (upper limits).

We test for the dependence of column density on the X-ray luminosity for the three *bona fide* CT samples with ATS method and found a marginal anticorrelation $\tau = -0.12$ with $P_{\rm ATS} = 0.24$ and $\tau = -0.12$ with $P_{\rm ATS} = 0.28$ for the *bona fide* CT Torus-selected sample and MYTorus-selected sample, respectively. This anticorrelation becomes slightly weaker for the case of the *bona fide* CT total sample ($\tau = -0.08$ with $P_{\rm ATS} = 0.38$). Those correlations are shown in Fig. 7. No correlation is found between the column density and the redshift $\tau = -0.09$ with $P_{\rm ATS} = 0.36$ and $\tau = -0.06$ with $P_{\rm ATS} = 0.50$ for the *bona fide* CT Torus-selected sample and the *bona fide* CT total sample, respectively. This correlation appears significant for the case of the *bona fide* CT MYTorus-selected sample ($\tau = -0.15$ with $P_{\rm ATS} = 0.20$) as shown in Fig. 8.

We investigated the presence of a correlation between the photon index and the column density or the X-ray luminosity but found no correlation in either of them for the *bona fide* CT Torus-selected and MYTorus-selected samples. Also, no correlation was detected for the *bona fide* CT total sample (see Fig. 9). Previously, an anticorrelation between the photon index and the X-ray luminosity was reported for type 1 AGN sample with $r_s = -0.21$ and $p = 8 \times 10^{-4}$ (Corral et al. 2011). Brightman et al. (2013) found a high significant correlation between $\Gamma$ and $\lambda_{\rm Edd}$ and confirmed that $\lambda_{\rm Edd}$ rather than $L_{\rm X}$, $L_{\rm UV}$, FWHM, or $M_{\rm BH}$ is the parameter which most strongly correlates with the X-ray spectral index, $\Gamma$. However, we do not have sufficient measurements of the Eddington ratio in our sample to validate this hypothesis. Finally, in the *bona fide* CT total sample there is no correlation between the photon index and redshift (see Fig. 10). Previously, a strong correlation was found for a sample of sources with $z > 2$ by Mateos et al. (2010) but most sources in our samples have $z < 2$.

We have detected five sources with significant neutral Fe K$\alpha$ emission line in the *bona fide* CT total sample. To check for possible dependence of the EW on luminosity, we used Spearman rank correlation analysis. We found a significant anticorrelation between the EW of the neutral Fe line and the X-ray luminosity at 2–

[2] https://CRAN.R-project.org/package=NADA
[3] https://github.com/jmeyers/linmix

**Table 3.** Spectral results of MYTorus fits to the *probable* CT sample (62 sources).

| Source (1) | $P_{CT}$ (2) | Γ (3) | $N_H$ ($10^{24}$cm$^{-2}$) (4) | $E$ (keV) (5) | $EW_{6.4}$ (keV) (6) | $\log_{10} L_X$ 2–10 keV (7) | Flux ($10^{-14}$) erg cm$^{-2}$ s$^{-1}$ (8) | C-stat/dof (9) |
|---|---|---|---|---|---|---|---|---|
| 3XMMJ085707.9+275319 | 0.62 | $2.35^{+0.10}_{-0.12}$ | $0.83^{+0.38}_{-0.40}$ | – | – | 45.21 | 7.35 | 259.23/310 |
| 3XMMJ090044.9+385521 | 0.0 | 2.21 ± 0.11 | $0.42^{+0.50}_{-0.19}$ | – | – | 44.58 | 10.18 | 247.06/356 |
| 3XMMJ095942.9+251524 | 0.70 | 2.33 ± 0.11 | $1.24^{+0.52}_{-0.60}$ | – | – | 44.57 | 8.75 | 386.48/425 |
| **3XMMJ100114.8+020208** | **1** | **>2.26** | **>5.23** | – | – | **46.10** | **10.38** | **372.63/364** |
| **3XMMJ100159.7+022641** | **0.96** | **2.08 ± 0.12** | $2.01^{+1.32}_{-0.66}$ | – | – | **46.85** | **7.40** | **497.14/535** |
| **3XMMJ102836.0+240402** | **0.98** | $2.18^{+0.13}_{-0.05}$ | $1.69^{+1.21}_{-0.85}$ | – | – | **46.00** | **19.88** | **303.13/387** |
| 3XMMJ112253.5+052025 | 0.0 | 2.04 ± 0.10 | $0.47^{+0.48}_{-0.29}$ | – | – | 44.48 | 11.92 | 376.86/462 |
| 3XMMJ104421.8+063545 | 0.0 | 2.38 ± 0.12 | $0.40^{+0.46}_{-0.24}$ | – | – | 45.53 | 3.32 | 375.91/379 |
| 3XMMJ115339.3+224920[c] | 0.23 | 2.44 ± 0.10 | $0.53^{+0.39}_{-0.21}$ | – | – | 46.09 | 6.85 | 478.62/527 |
| **3XMMJ120049.8+341852** | **1** | $2.11^{+0.13}_{-0.09}$ | **>2.94** | – | – | **45.26** | **3.01** | **379.60/440** |
| **3XMMJ121844.2+055714**[b] | **0.97** | **1.80 ± 0.10** | $2.52^{+0.57}_{-0.63}$ | – | – | **44.64** | **8.55** | **431.55/509** |
| **3XMMJ080622.9+152731**[b] | **0.99** | **2.6** | $1.91^{+0.38}_{-0.30}$ | – | – | **47.03** | **4.56** | **1018.96/1069** |
| **3XMMJ125331.9+100532**[b] | **1** | **2.12 ± 0.09** | $1.04^{+1.08}_{-0.54}$ | – | – | **45.69** | **13.17** | **508.52/530** |
| **3XMMJ132930.3+114857** | **0.99** | **2.22 ± 0.12** | $1.33^{+5.02}_{-0.59}$ | – | – | **45.48** | **6.29** | **396.93/437** |
| **3XMMJ140638.2+222117** | **1** | $1.78^{+0.09}_{-0.07}$ | **>2.43** | **6.35 ± 0.04** | $0.39^{+0.00}_{-0.005}$ | **45.64** | **15.04** | **467.81/585** |
| 3XMMJ140858.3+261239[c] | 0.0 | 2.31 ± 0.06 | $0.80^{+0.32}_{-0.24}$ | – | – | 45.55 | 7.63 | 916.59/925 |
| **3XMMJ011829.6+004548** | **0.85** | **2.21 ± 0.07** | $1.03^{+2.86}_{-0.67}$ | – | – | **45.16** | **24.25** | **611.78/688** |
| 3XMMJ083734.6+060644[b] | 0.54 | 2.09 ± 0.12 | $0.78^{+0.94}_{-0.46}$ | – | – | 45.70 | 3.10 | 407.39/445 |
| **3XMMJ083827.1+255050** | **0.95** | $1.94^{+0.22}_{-0.18}$ | **>3.67** | – | – | **46.64** | **39.69** | **302.00/392** |
| **3XMMJ100056.5+554059**[b] | **0.91** | **1.76 ± 0.12** | $1.37^{+1.27}_{-0.65}$ | – | – | **46.64** | **18.43** | **353.64/386** |
| 3XMMJ100101.5+250904 | 0.0 | $2.06^{+0.34}_{-0.30}$ | <0.78 | 6.37 ± 0.06 | $0.44^{+0.002}_{-0.005}$ | 45.34 | 3.14 | 375.71/473 |
| **3XMMJ100336.5+325154**[b] | **0.89** | $1.99^{+0.32}_{-0.08}$ | **1.28** | – | – | **44.13** | **6.59** | **423.18/511** |
| **3XMMJ102224.2+383504** | **1** | **1.95 ± 0.07** | **>0.36** | – | – | **44.25** | **15.67** | **397.37/461** |
| **3XMMJ102408.3+040931** | **0.96** | **1.55** | $1.50^{+1.16}_{-0.41}$ | – | – | **45.32** | **1.95** | **122.44/129** |
| **3XMMJ103446.8+575128** | **1** | **2.25 ± 0.16** | **>4.41** | – | – | **45.81** | **5.89** | **405.51/502** |
| **3XMMJ104048.5+061819**[b] | **0.90** | **1.87 ± 0.05** | $1.37^{+0.44}_{-0.50}$ | – | – | **45.66** | **44.13** | **814.00/859** |
| 3XMMJ111238.1+132245[b] | 0.0 | $2.09^{+0.13}_{-0.26}$ | 0.02 ± 0.01 | 6.42 ± 0.05 | 0.50 ± 0.002 | 45.50 | 7.04 | 451.30/477 |
| **3XMMJ111247.5+132419**[b] | **0.97** | **1.65 ± 0.08** | **1.50** | – | – | **44.10** | **3.70** | **432.71/541** |
| **3XMMJ113322.8+663326** | **1** | **2.29 ± 0.15** | **>2.50** | – | – | **46.39** | **29.77** | **595.63/624** |
| **3XMMJ115010.2+550709**[b] | **0.92** | **1.87 ± 0.10** | $1.33^{+1.61}_{-0.66}$ | – | – | **47.02** | **9.35** | **442.09/569** |
| **3XMMJ121047.2+392313** | **1** | $1.90^{+0.18}_{-0.15}$ | **>4.56** | – | – | **46.05** | **4.44** | **504.38/559** |
| 3XMMJ121051.2+393206[b] | 0.62 | 1.95 ± 0.24 | $1.15^{+0.25}_{-0.22}$ | – | – | 46.40 | 7.02 | 244.10/271 |
| 3XMMJ121955.1+054643 | 0.0 | $2.35^{+0.20}_{-0.22}$ | <0.02 | 6.40 ± 0.03 | $0.63^{+0.001}_{-0.005}$ | 45.41 | 2.27 | 560.30/668 |
| **3XMMJ122935.6+075326** | **0.99** | **2.00 ± 0.10** | **2.16** | – | – | **45.19** | **1.92** | **420.03/443** |
| **3XMMJ123047.0+105606** | **1** | **>2.48** | **>2.79** | – | – | **45.45** | **4.78** | **558.46/592** |
| **3XMMJ123121.4+135955**[b] | **1** | **2.00 ± 0.12** | **>0.47** | – | – | **46.01** | **4679** | **309.96/362** |
| **3XMMJ123140.5+254342** | **0.98** | $1.65^{+0.14}_{-0.07}$ | $7.07^{+1.20}_{-1.46}$ | – | – | **45.33** | **10.39** | **363.92/439** |
| **3XMMJ130307.7−022427** | **1** | **>2.09** | **>6.54** | – | – | **46.60** | **12.31** | **347.68/436** |
| **3XMMJ132529.9+320117**[b] | **0.95** | $1.67^{+0.08}_{-0.05}$ | $1.71^{+3.76}_{-0.88}$ | – | – | **46.06** | **30.20** | **504.33/597** |
| **3XMMJ134859.4+601457** | **1** | **2.13 ± 0.12** | **>2.38** | – | – | **45.40** | **7.37** | **428.26/480** |
| **3XMMJ140726.5+524710** | **1** | **>2.16** | **>4.41** | – | – | **46.37** | **11.63** | **334.48/402** |
| **3XMMJ143031.3–000907** | **1** | $2.13^{+0.18}_{-0.20}$ | $1.69^{+0.65}_{-0.47}$ | – | – | **47.25** | **8.42** | **328.10/406** |
| **3XMMJ144250.0−004249** | **0.97** | **2.09 ± 0.13** | $1.14^{+1.50}_{-0.56}$ | $6.31^{+0.07}_{-0.05}$ | **0.20 ± 0.0** | **46.43** | **4.96** | **380.27/467** |
| **3XMMJ143013.8−001020** | **0.98** | $2.00^{+0.10}_{-0.06}$ | $2.66^{+1.28}_{-0.77}$ | – | – | **45.16** | **39.72** | **415.28/444** |
| **3XMMJ151640.1−010212**[b] | **0.92** | **1.64 ± 0.06** | **1.64 ± 0.83** | – | – | **44.70** | **13.26** | **634.42/693** |
| 3XMMJ130336.8+673027 | 0.0 | 2.23 ± 0.06 | $0.37^{+1.51}_{-0.20}$ | – | – | 45.26 | 17.12 | 805.71/889 |
| **3XMMJ153248.5+043 948** | **1** | **2.16 ± 0.15** | **>5.19** | – | – | **46.13** | **6.61** | **396.61/496** |
| **3XMMJ224304.5−094241**[b] | **0.95** | **2.06 ± 0.08** | **>0.27** | – | – | **44.20** | **2.89** | **618.64/795** |
| 3XMMJ234715.9+005602[b] | 0.06 | 2.34 ± 0.09 | $0.91^{+0.54}_{-0.42}$ | – | – | 45.61 | 7.91 | 484.87/504 |
| **3XMMJ133654.6+514611** | **0.97** | $2.11^{+0.13}_{-0.11}$ | **>4.10** | – | – | **45.75** | **11.41** | **465.09/552** |
| **3XMMJ111715.8+175741** | **1** | **1.9**[a] | **>8.92** | – | – | **46.24** | **35.61** | **707.16/782** |

**Table 3** – *continued*

| Source (1) | $P_{CT}$ (2) | $\Gamma$ (3) | $N_H$ ($10^{24}$cm$^{-2}$) (4) | $E$ (keV) (5) | $EW_{6.4}$ (keV) (6) | $\log_{10} L_X$ 2–10 keV (7) | Flux ($10^{-14}$) erg cm$^{-2}$ s$^{-1}$ (8) | C-stat/dof (9) |
|---|---|---|---|---|---|---|---|---|
| 3XMMJ115903.3+440149[b] | 0.40 | $1.65^{+0.90}_{-0.25}$ | $0.06 \pm 0.04$ | – | – | 44.22 | 1.87 | 73.93/83 |
| 3XMMJ121010.7+392300[b] | 0.0 | $2.06 \pm 0.50$ | $0.39^{+0.13}_{-0.11}$ | – | – | 44.05 | 2.33 | 88.19/95 |
| 3XMMJ122558.8+332823[b] | 0.04 | $2.16 \pm 0.43$ | $0.37^{+0.14}_{-0.10}$ | – | – | 44.38 | 1.43 | 139.09/128 |
| **3XMMJ151616.1+000148** | **0.97** | **2.6** | $3.76^{+1.31}_{-0.84}$ | – | – | **45.50** | **8.47** | **80.63/96** |
| 3XMMJ105106.3+573435 | 0.0 | $2.36^{+0.23}_{-0.96}$ | $0.16 \pm 0.07$ | – | – | 45.12 | 3.29 | 99.19/111 |
| 3XMMJ030705.8−000009[b] | 0.0 | 2.6 | $0.48^{+0.28}_{-0.20}$ | – | – | 44.82 | 4.61 | 535.99/569 |
| 3XMMJ093535.4+611919 | 0.0 | $2.39 \pm 0.07$ | $0.54^{+1.95}_{-0.28}$ | – | – | 44.69 | 6.20 | 625.31/713 |
| 3XMMJ123601.9+263509[b] | 0.44 | $2.08 \pm 0.12$ | $0.23^{+0.19}_{-0.13}$ | – | – | 45.60 | 4.90 | 400.19/409 |
| 3XMMJ131240.8+230924 | 0.0 | $2.36 \pm 0.07$ | $0.33^{+0.27}_{-0.14}$ | – | – | 45.35 | 8.40 | 732.65/783 |
| **3XMMJ085846.5+274534** | **1** | $\mathbf{2.29 \pm 0.10}$ | $3.41^{+0.80}_{-0.67}$ | – | – | **44.97** | **6.79** | **643.90/732** |
| 3XMMJ140921.7+261317 | 0.0 | 1.9[a] | $0.08 \pm 0.02$ | – | – | 45.72 | 5.11 | 244.68/223 |

Columns: (1) Sources name; (2) Probability of being CT from MYTorus model; (3) Photon index; (4) Column density; (5) Centre energy of the 6.4 keV line; (6) EW of the 6.4 keV line; (7) The logarithm of the inferred luminosity in the 2–10 keV band; (8) 2–10 keV absorbed flux; (9) C-stat to the numbers of degrees of freedom. All errors are at 1$\sigma$ confidence level except the EW for the 6.4 keV line is 90 per cent confidence level.
[a]fixed parameter;
[b]$A_S$ not constrained;
[c]$A_z$ is zero. Sources in bold are the *bona fide* CT MYTorus-selected sample.

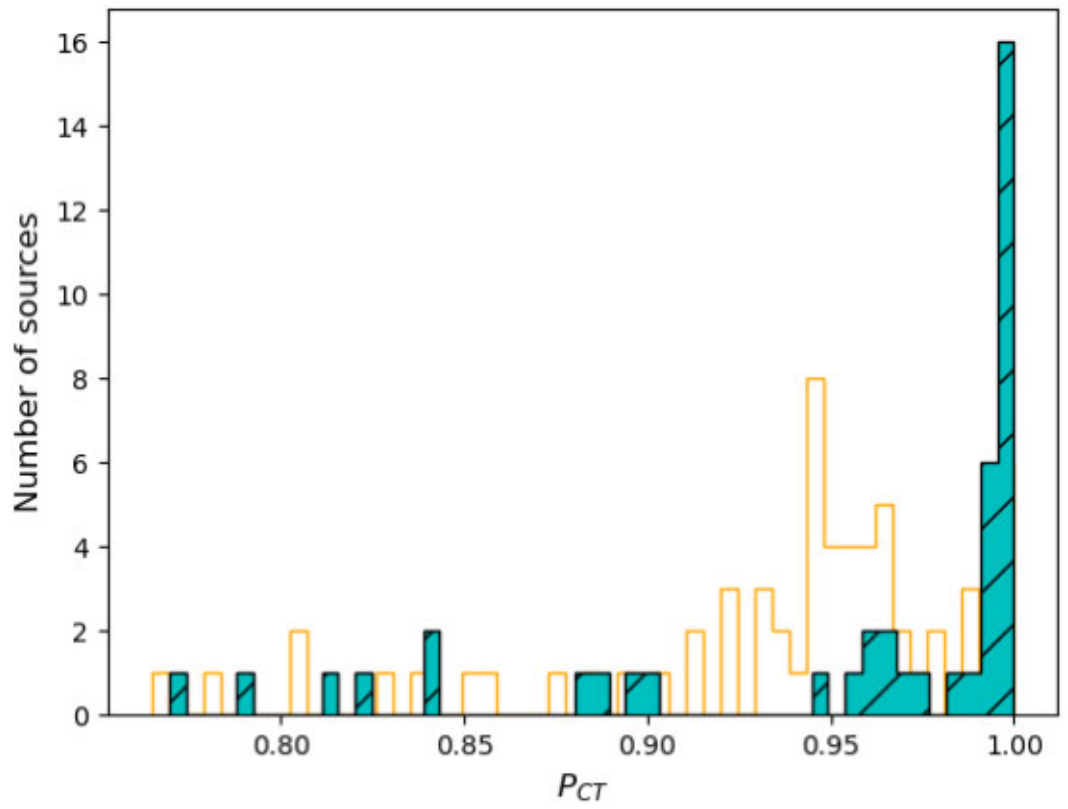

**Figure 4.** The distribution of $P_{CT}$ for the *bona fide* CT Torus-selected sample (solid histogram) and the *bona fide* CT MYTorus-selected sample (shaded histogram).

10 keV ($r_s = -0.70$ with $p = 0.19$). To quantify this correlation, we performed linear regression using LINMIX-ERR package. The anticorrelation between the neutral line EW and 2–10 keV X-ray luminosity as well as the best-fitting line and 1$\sigma$ confidence interval are plotted in Fig. 11. The best-fitting line is given as:

$$\text{Log } EW = (3.21 \pm 10.02) + (-0.36 \pm 5.87) \log L_{X,44},$$

where $EW$ is the equivalent width of the neutral iron K$\alpha$ line in eV and $L_{X,44}$ is the 2–10 keV X-ray luminosity in units of $10^{44}$ erg s$^{-1}$. We find a correlation coefficient of $-0.75^{+0.36}_{-0.16}$. This shows a significant anticorrelation, consistent with the Spearman rank test. This correlation agrees with the X-ray Baldwin effect. A common interpretation of this effect is related to the covering factor of the torus (Zhou & Wang 2005) which is proportional to the EW of the iron K$\alpha$ line (Ikeda, Awaki & Terashima 2009) and therefore the covering factor of the torus decreases with luminosity. For distant galaxies $0.5 < z < 2$, it is believed that the observed iron K$\alpha$ line is extended to galactic scale (kpc), beyond the parsec-scale torus, due to reflection caused by galactic star-forming clouds (Fabbiano et al. 2017; Jones et al. 2020; Wei Yan et al. 2021; Gilli et al. 2022). Because of the low number of sources examined (only five sources) the coefficients of the best-fitting line are inaccurate. Consequently, we could not compare our slope with the previous works: Page et al. (2004; $-0.17 \pm 0.08$) for type-I AGN, Shu et al. (2010; $-0.13 \pm 0.04$) for unobscured and moderately obscured AGN and Shu et al. (2012; $-0.11 \pm 0.03$) for unobscured AGN.

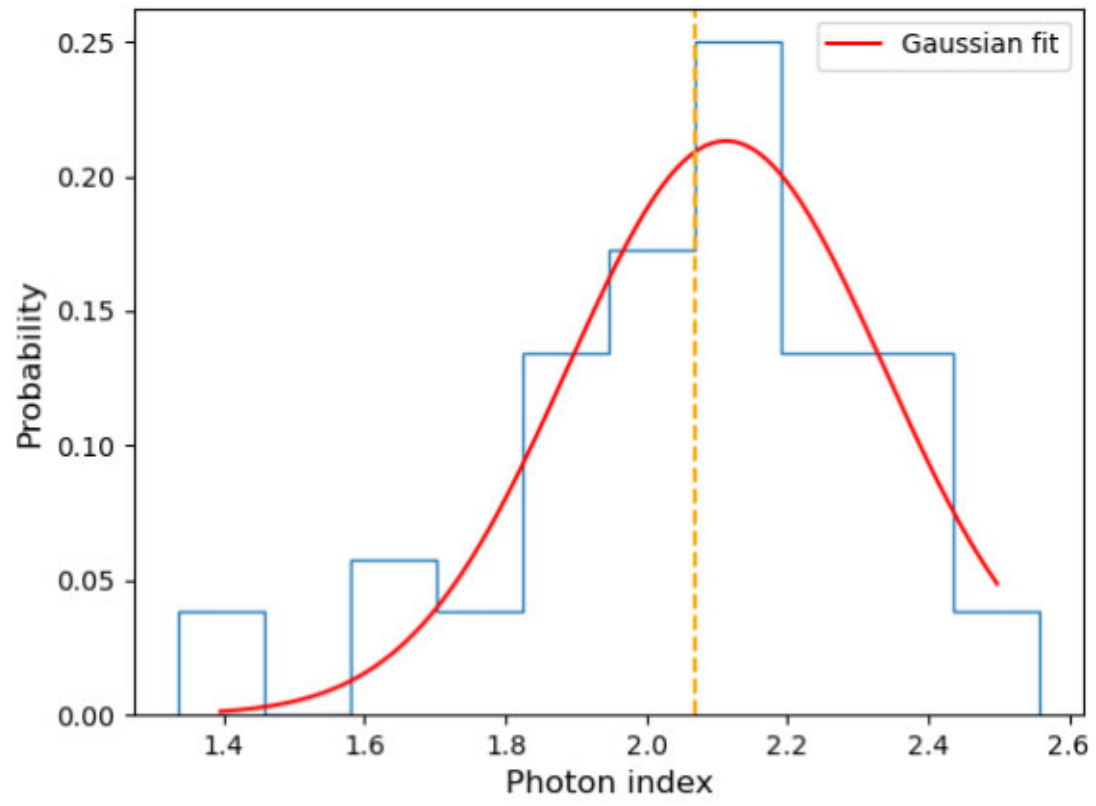

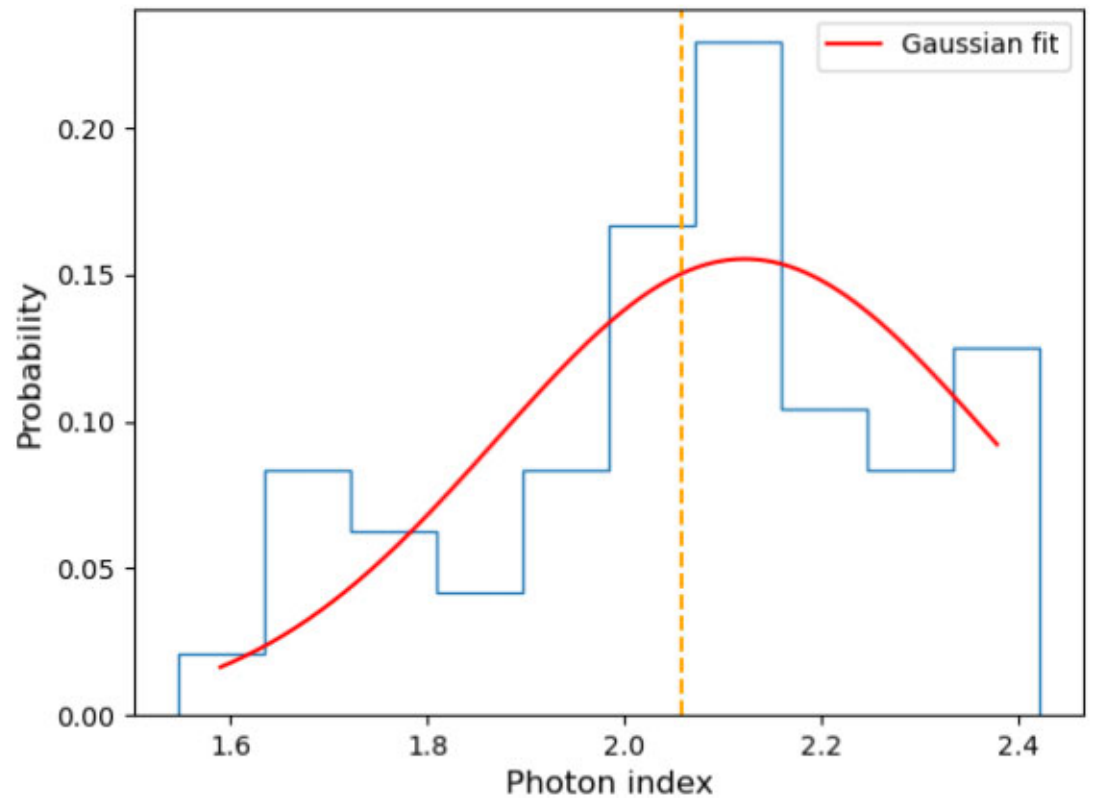

**Figure 5.** Distribution of the photon index $\Gamma$ for the *bona fide* CT total sample fitted with Gaussian. The solid vertical line refers to the mean $\Gamma$. Photon index distribution obtained from fitting with Torus model is on top and from fitting with MYTorus model is in bottom.

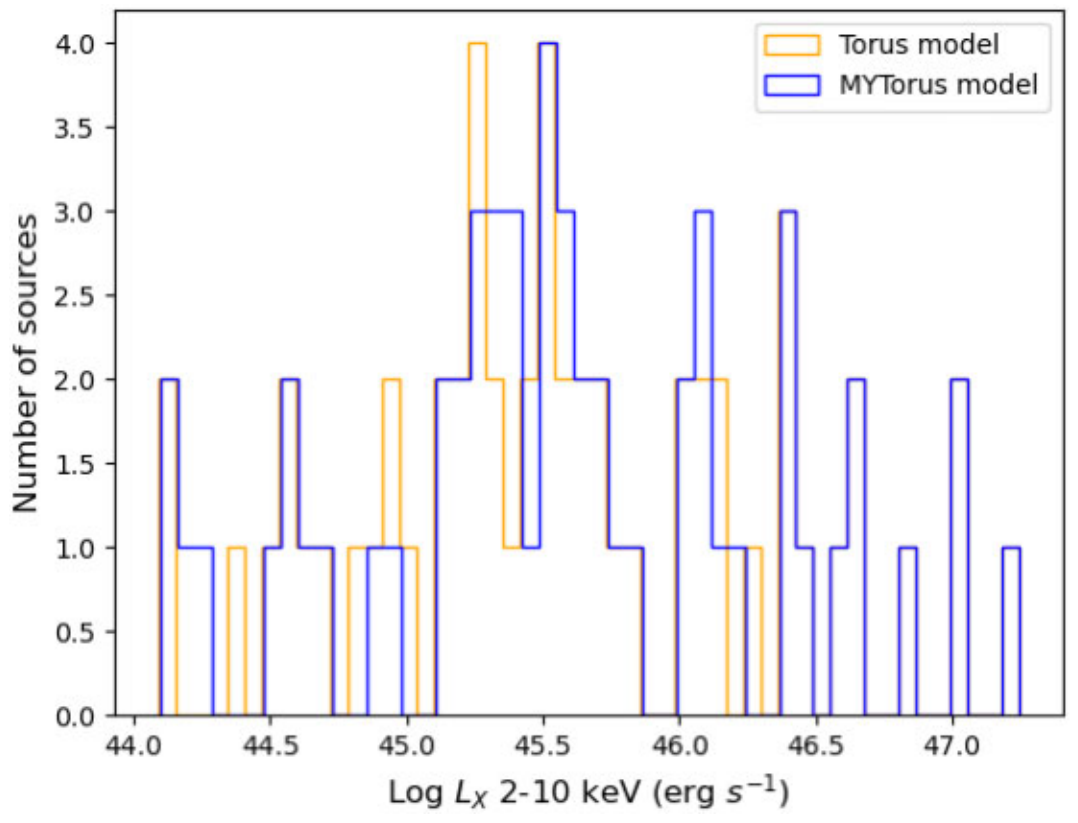

**Figure 6.** Hard X-ray (2–10 keV) luminosity for the *bona fide* CT total sample.

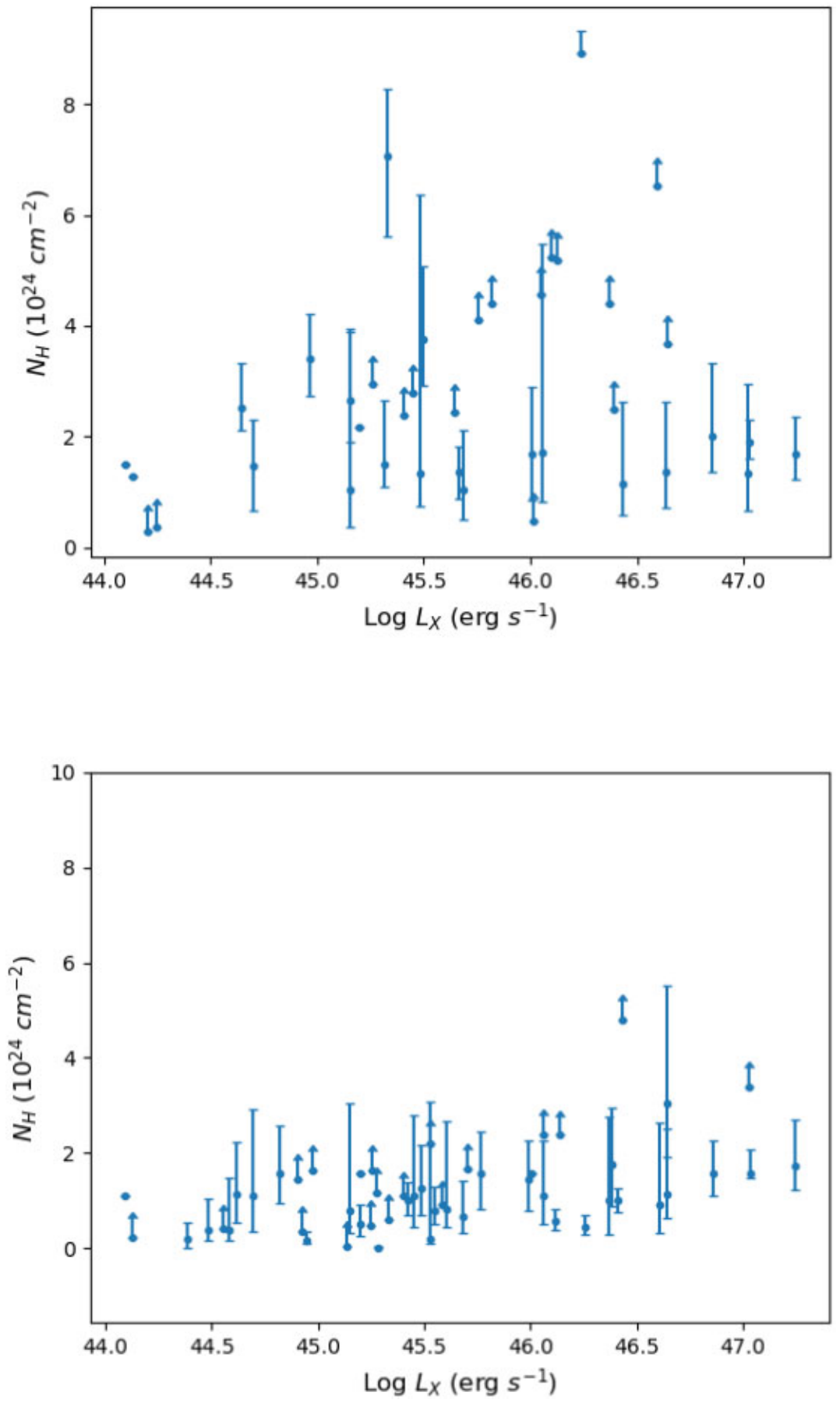

**Figure 7.** Plot of the column density and X-ray luminosity (2–10 keV) for the *bona fide* CT MYTorus-selected sample (top) and the *bona fide* CT total sample (bottom). Arrows correspond to lower limits on the column density.

We do not find a correlation between the column density and the EW of the 6.4 KeV line in the *bona fide* CT total sample ($\tau = -0.33$ with $P_{\rm ATS} = 0.64$). A positive correlation between the column density and EW was reported (Guainazzi, Matt & Perola 2005; Fukazawa et al. 2011) for sources with $N_{\rm H}$ above $10^{23}$ cm$^{-2}$. A similar result was found in a sample of highly absorbed AGN by using *XMM–Newton* data (Corral et al. 2014).

## 5 DISCUSSION

In this work, we select a sample of *bona fide* CT AGN composed of 52 sources. Our selection of CT AGN is based on a newly used selection technique based on an MCMC. The advantage of this selection technique is that it does not rely on the best-fitting value of column density in the phenomenological models but instead it estimates through simulations the probability of being CT AGN based on physically motivated models. We investigated the goodness of MCMC method of selection for CT AGN using Monte Carlo simulations. For the *bona fide* CT total sample, we selected CT candidates when either of the two models (Torus or MYTorus model) yields a $P_{\rm CT}$ ranges from 0.75 to 1 (see sources in bold in Tables 2 and 3). About 75 per cent of CT candidates in our sample have a very high probability ($P_{\rm CT} > 0.9$) of being CT.

Out of 52 CT candidates in our sample, about 80 per cent (42 sources) belong to HEW subsample and about 16 per cent (9 sources)

**Table 4.** Summary of the correlations found in our *bona fide* CT Torus-selected, MYTorus-selected, and total samples.

| Correlation (1) | Torus-selected sample (2) | MYTorus-selected sample (3) | Total sample (Torus) (4) |
|---|---|---|---|
| $N_{\rm H}$ and $L_{\rm X}$ (ATS) | $\tau = -0.12$, $P_{\rm ATS} = 0.24$ | $\tau = -0.12$, $P_{\rm ATS} = 0.28$ | $\tau = -0.08$, $P_{\rm ATS} = 0.38$ |
| $N_{\rm H}$ and flux (ATS) | $\tau = -0.02$, $P_{\rm ATS} = 0.83$ | $\tau = -0.03$, $P_{\rm ATS} = 0.81$ | $\tau = -0.03$, $P_{\rm ATS} = 0.77$ |
| $N_{\rm H}$ and z (ATS) | $\tau = -0.09$, $P_{\rm ATS} = 0.36$ | $\tau = -0.15$, $P_{\rm ATS} = 0.20$ | $\tau = -0.06$, $P_{\rm ATS} = 0.50$ |
| $N_{\rm H}$ and $\Gamma$ (ATS) | $\tau = 0.0$, $P_{\rm ATS} = 1$ | $\tau = -0.01$, $P_{\rm ATS} = 0.96$ | $\tau = 0.02$, $P_{\rm ATS} = 0.85$ |
| $L_{\rm X}$ and $\Gamma$ (ATS) | $\tau = 0.0$, $P_{\rm ATS} = 1$ | $\tau = -0.01$, $P_{\rm ATS} = 0.94$ | $\tau = 0.0$, $P_{\rm ATS} = 1$ |
| z and $L_{\rm X}$ (Spearman's rank) | $r_{\rm s} = 0.90$, $p = 0.0$ | $r_{\rm s} = 0.82$, $p = 0.0$ | $r_{\rm s} = 0.89$, $p = 0.0$ |
| Flux and $L_{\rm X}$ (Spearman's rank) | $r_{\rm s} = 0.18$, $p = 0.23$ | $r_{\rm s} = 0.29$, $p = 0.08$ | $r_{\rm s} = 0.20$, $p = 0.14$ |
| z and flux (Spearman's rank) | $r_{\rm s} = -0.18$, $p = 0.23$ | $r_{\rm s} = -0.18$, $p = 0.30$ | $r_{\rm s} = -0.17$, $p = 0.21$ |

**Notes.** Columns: (1) Parameters of the correlation (correlation test used); (2) The *bona fide* CT Torus-selected sample; (3) The *bona fide* CT MYTorus-selected sample; (4) The *bona fide* CT total sample with parameters from the Torus model. $r_{\rm s}$ and $\tau$ are the correlation coefficients for Spearman's Rank and ATS, respectively. $p$ and $P_{\rm ATS}$ are the probability of the correlation for Spearman's rank and ATS, respectively.

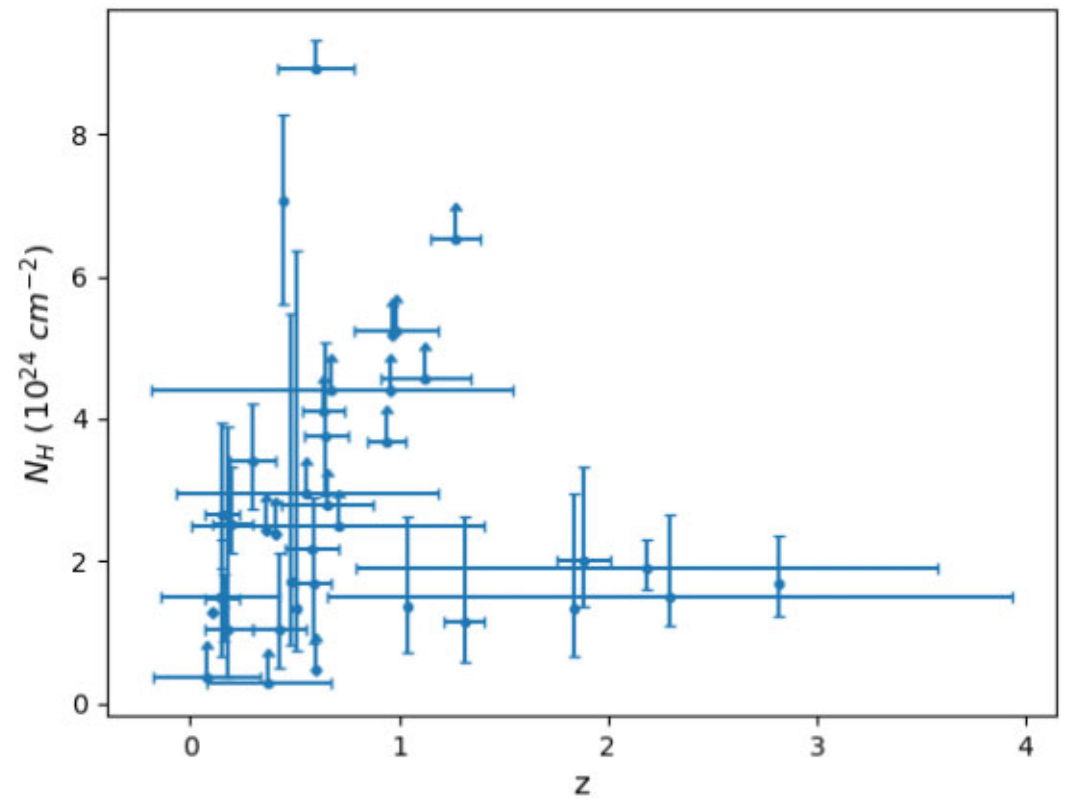

**Figure 8.** Column density versus redshift for the *bona fide* CT MYTorus-selected sample. Arrows correspond to lower limits on the column density.

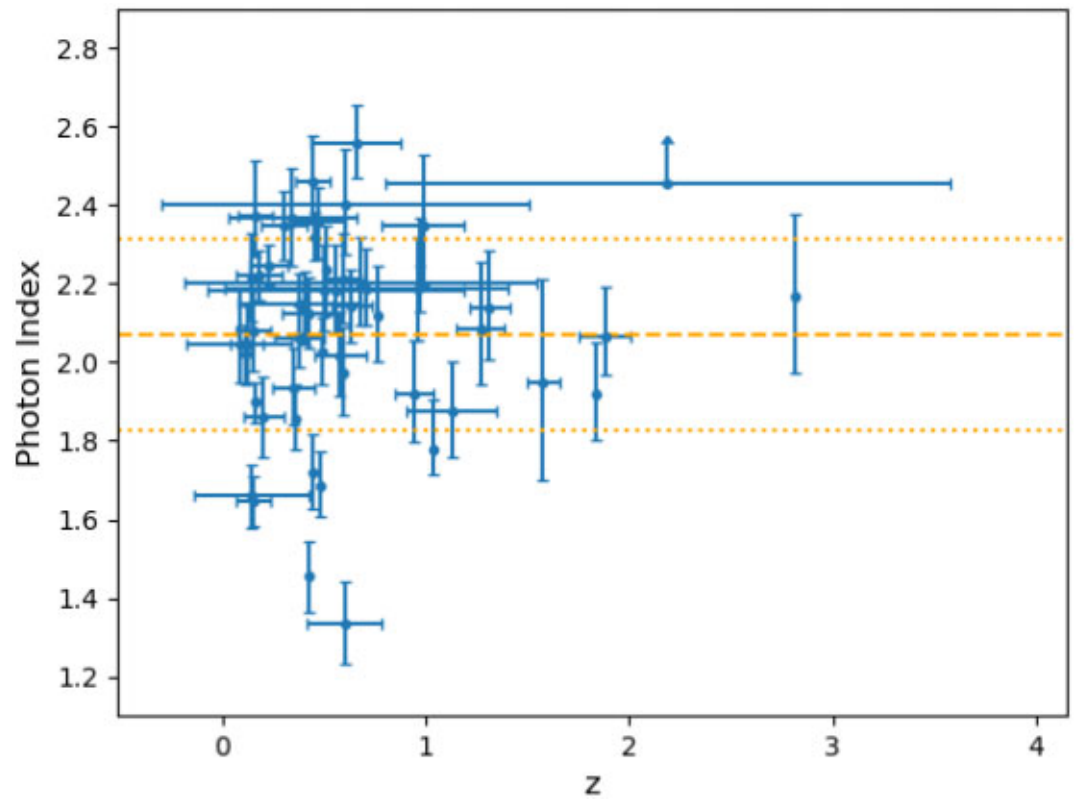

**Figure 10.** Redshifts versus Γ for the *bona fide* CT total sample. The dashed and dotted horizontal lines represent the mean value of Γ and its deviation, respectively.

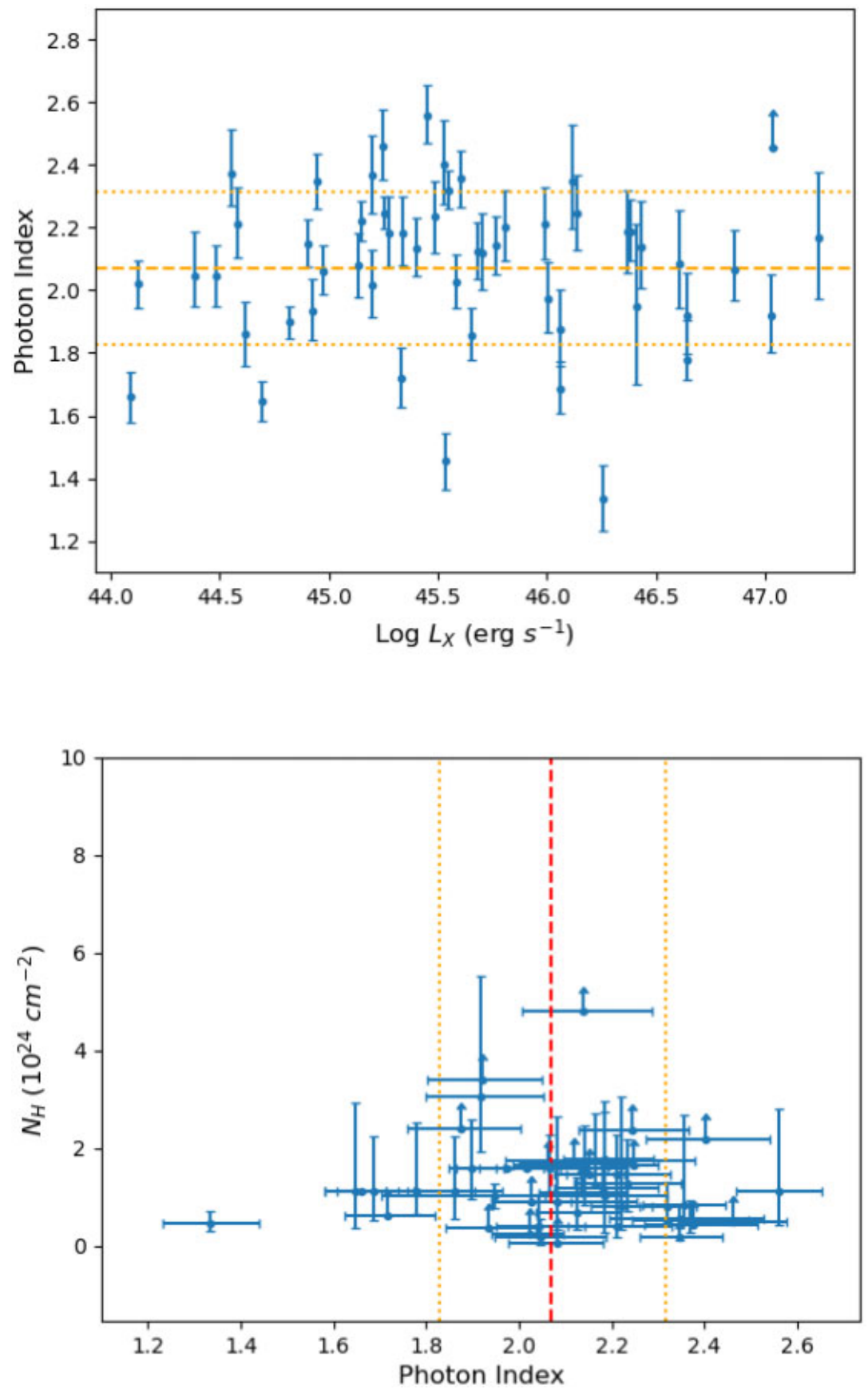

**Figure 9.** Γ versus the hard X-ray luminosities (top) and the column density (bottom) for the *bona fide* CT total sample. The dashed and dotted lines represent the mean value of Γ and its deviation, respectively.

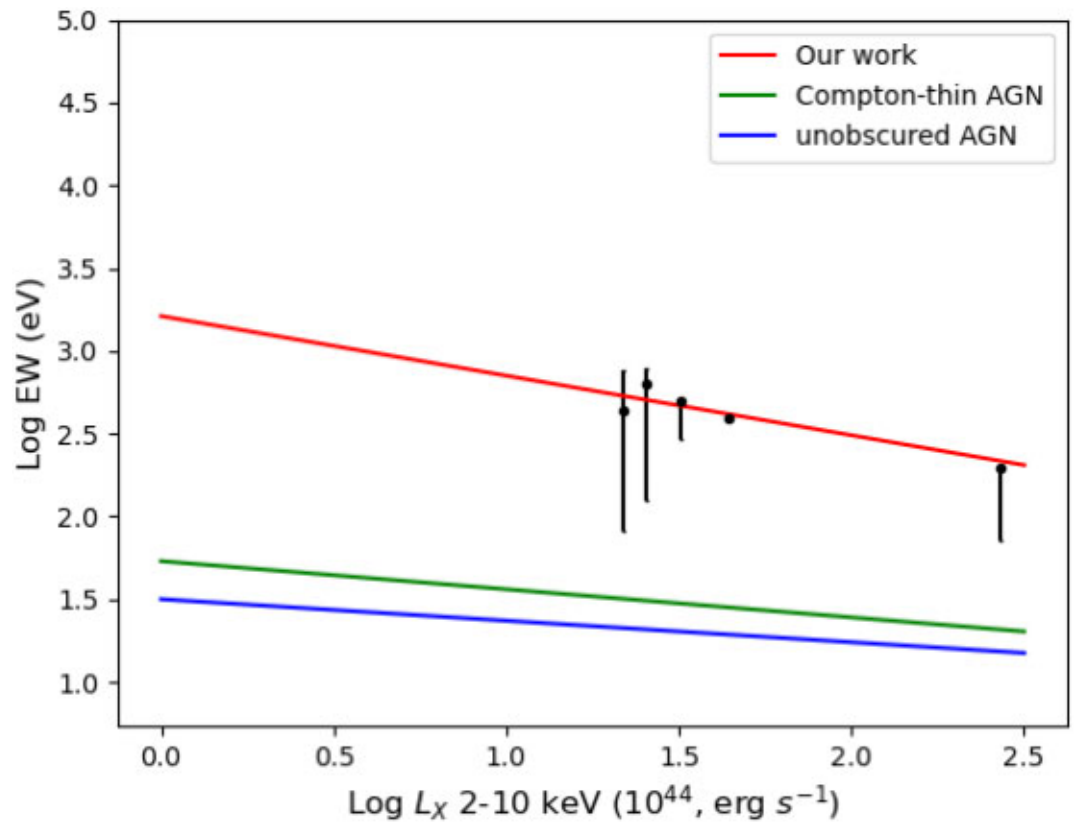

**Figure 11.** Fe K$\alpha$ EW versus 2–10 keV X-ray luminosity. The red line represents the best-fitting line of our sources. The green line shows the Iwasawa–Taniguchi relation for Compton-thin objects (Bianchi et al. 2007) (−0.17 ± 0.03). The blue line shows the Iwasawa–Taniguchi relation for unobscured and moderately obscured AGN (Shu et al. 2010; −0.13 ± 0.04).

belong to FLATH subsample. Therefore, HEW is the best criterion subsample to search for CT candidates. We searched in the literature for previous works that classified any source of our sample as CT. Zezas et al. (2005) presented results from a *Chandra* ACIS-S observation of the type 2 low-ionization nuclear emission-line region radio galaxy NGC 4261 and the X-ray data showed that this galaxy hosts a heavily obscured AGN. NGC 2623 was previously detected as an obscured AGN in Maiolino et al. (2003). By using *Chandra*/ACIS-S data they showed evidence for a flat hard X-ray spectrum which they explained as the presence of a CT, cold reflection-dominated AGN. Evidence of a flat spectrum was later presented by the analysis of optical *Hubble Space Telescope*, *Spitzer Space Telescope*, and *XMM* observations (Evans et al. 2008). For the high probable CT source IRAS F12514+1027 ($P_{\rm CT} = 0.88$ and 1 from Torus and MYTorus model, respectively) in our CT sample, Wilman et al. (2003) reported that in the hard X-ray band ($E > 2$ keV) the flat continuum and Fe K absorption edge suggest that the emission from the active nucleus is reflection-dominated, possibly implying that the direct line of sight to the nucleus is obscured by CT material.

Recently, Ruiz et al. (2021) have provided a sample of 977 X-ray absorbed AGN from the XMMFITCAT-Z catalogue. We searched for common sources with their sample, but we did not find common sources since their sample only includes 19 CT sources (out of 977 absorbed sources). Additionally, they noted that CT sources might not have been detected in their final sample of X-ray absorbed AGNs because iron emission lines were not included in the models used by XMMFITCAT-Z to look for absorbed sources, while the majority of our CT sources were found in the HEW sample.

It has been suggested that besides the absorption from the parsec-scale torus there exists an absorption at subparsec distances from the central source, mostly from the BLR clouds (Bianchi et al. 2012). The presence of such absorption is detected through absorption variability measurements. Changes in the line-of-sight column density were detected for several AGNs such as NGC 6300 (Guainazzi 2002), UGC 4203 (Guainazzi et al. 2002), and NGC 5506 (Markowitz, Krumpe & Nikutta 2014). Sometimes AGNs change from Compton-thin to CT or vice versa, so-called changing-look AGNs (e.g. IC 751 Ricci et al. 2016). These absorption variations could happen in short time-scale such as in the case of NGC 1365 which changes from CT to CT on time-scales from a couple of days to ∼10 h (Risaliti et al. 2005). One probable CT AGNs selected by our analysis, NGC 4395 was found previously to have variations in the $N_{\rm H}$ (Nardini & Risaliti 2011) but have never been previously reported as changing-look AGNs.

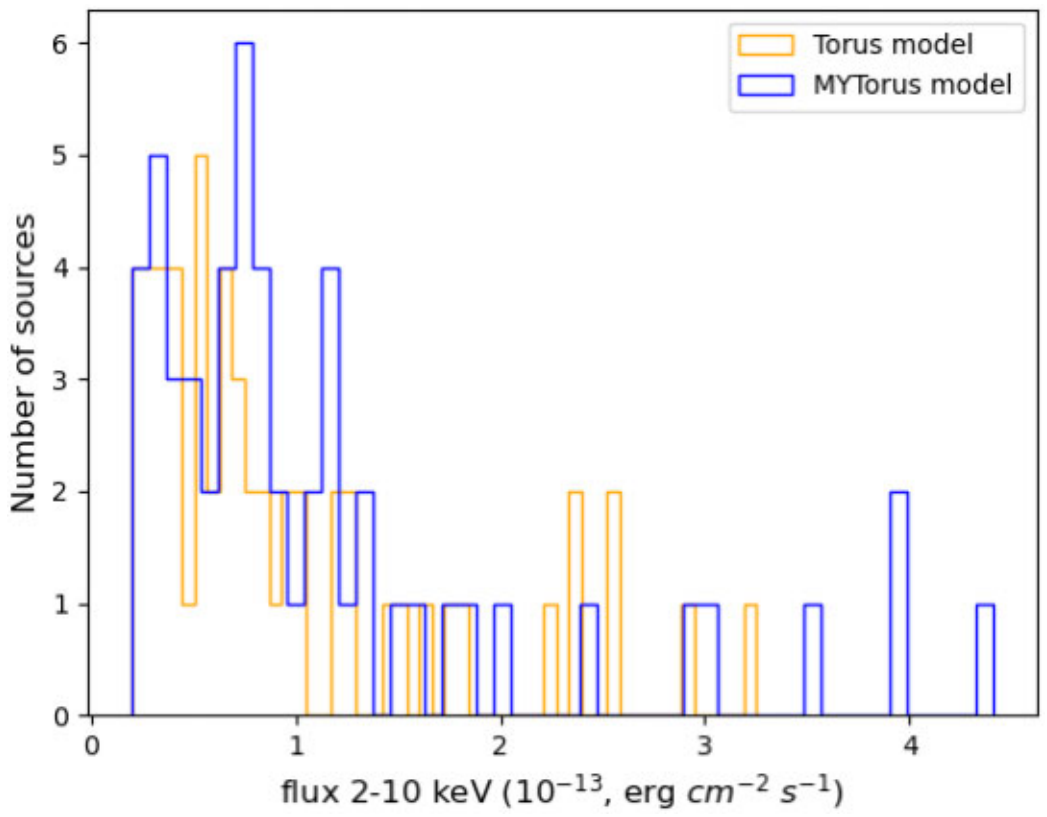

**Figure 12.** 2–10 keV absorbed flux for the *bona fide* CT total sample.

All CT sources in our *bona fide* CT total sample have absorbed 2–10 keV flux below $4.5 \times 10^{-13}$ erg cm$^{-2}$ s$^{-1}$ as indicated in their flux distribution in Fig. 12. Similarly, Guainazzi et al. (2005) observed that the CT sources largely dominate the sample below $5 \times 10^{-13}$ erg cm$^{-2}$ s$^{-1}$. This is consistent with the fact that CT sources are more frequently discovered at fainter fluxes (see Malizia et al. 2009). This is as a result of the substantial obscuration that lowers the intrinsic flux.

We find a mean value of $\Gamma = 2.07 \pm 0.11$ with a deviation of $\sigma = 0.24$ and $\Gamma = 2.06 \pm 0.12$ with a deviation of $\sigma = 0.21$ of the *bona fide* CT total sample when fitting with Torus model and MYTorus model, respectively. It is rather steeper than the mean value $\Gamma = 1.95 \pm 0.05$ with a dispersion of $\sigma = 0.15$ derived from the X-ray spectra of 27 Seyfert galaxies observed with *Ginga* LAC when the effects of reflection and warm absorber have been accounted for (Nandra & Pounds 1994). This could be interpreted through the fact that Seyfert 2 s spectra are harder than those of Seyfert 1 s. Zdziarsk, Poutanen & Johnson (2000) reported a significant difference in the weighted average of the indices of individual objects and the indices in the power-law fits to the sum spectra between Seyfert 1 s, $2.50 \pm 0.09$, and $2.56 \pm 0.14$, respectively, and Seyfert 2 s, $2.05 \pm 0.09$, and $2.21 \pm 0.1$. Middleton, Done & Schurch (2008) presented this finding more recently, stating that type 1 Seyfert galaxies having $\Gamma \sim 2.1$ whereas type 2 Seyfert galaxies are harder with $\Gamma \sim 1.9$. X-ray spectral analysis of X-ray bright sample composed of 82 sources in the 1Ms catalogue of the Chandra Deep Field South presented a mean value for the $\Gamma$ of $1.75 \pm 0.02$, with intrinsic dispersion of $\sigma \sim 0.30$ (Tozzi et al. 2006). The mean $\Gamma$ appears more flattened than our mean value of $\Gamma$. Molina et al. 2013 presented the hard X-ray spectral analysis of a sample of 41 type 1 AGN and 33 type 2 AGN and found mean value $\Gamma = 2.00 \pm 0.03$ with $\sigma = 0.19$ and mean value $\Gamma = 2.10 \pm 0.08$ with $\sigma = 0.41$, respectively. This is in contrary to Burlon et al. (2011), who used *Swift*/BAT data of 199 AGN and obtained mean value $\Gamma = 2.07 \pm 0.03$ with $\sigma = 0.27$ for unobscured sources and mean value $\Gamma = 1.92 \pm 0.02$ with $\sigma = 0.25$ for obscured objects.

The observed anticorrelation between the EW of neutral Fe K$\alpha$ emission line and the X-ray 2–10 keV luminosity found in our CT sample is consistent with the Iwasawa–Taniguchi effect. According to the Iwasawa–Taniguchi effect, high-luminosity AGNs have a smaller EW for the 6.4 keV line. This explains the lack of CT sources for higher luminosity AGN. Strong iron line features in CT AGNs can be explained by reflection from obscuring material in a small scale (∼1–100 pc) region (Ricci et al. 2014). However, recent works of low-redshift CT AGNs (Fabbiano et al. 2017; Jones et al. 2020) detected kpc-scale diffuse emission of the iron line from additional reflection of the X-ray emission extending to galactic scales rather than the parsec-scale torus. Similar results (Circosta et al. 2019; D'Amato et al 2020) also suggest that the AGN obscuration along the line of sight can be due to a contribution from the gas and dust in the host galaxy, instead of only the contribution from the torus region. Most recent work by Wei Yan et al. (2021) find a clear connection between far-IR luminosity (galactic star formation rate SFR) and the strength of the Fe K$\alpha$ line and they interpret this relation as being due to reflection of nuclear emission by the star-forming gas that is typically distributed on galaxy scales.

## ACKNOWLEDGEMENTS

We acknowledge XMMFITCAT-Z data base, based on results from the enhanced *XMM–Newton* spectral-fit data base, an ESA PRODEX funded project, based in turn on observations obtained with *XMM–Newton*, an ESA science mission with instruments and contributions directly funded by ESA Member States and NASA.

## DATA AVAILABILITY

The data underlying this article are available in the article and its online supplementary material.

## SUPPORTING INFORMATION

Supplementary data are available at *MNRAS* online.

**parentsample.fits**

Please note: Oxford University Press is not responsible for the content or functionality of any supporting materials supplied by the authors. Any queries (other than missing material) should be directed to the corresponding author for the article.

**APPENDIX A**

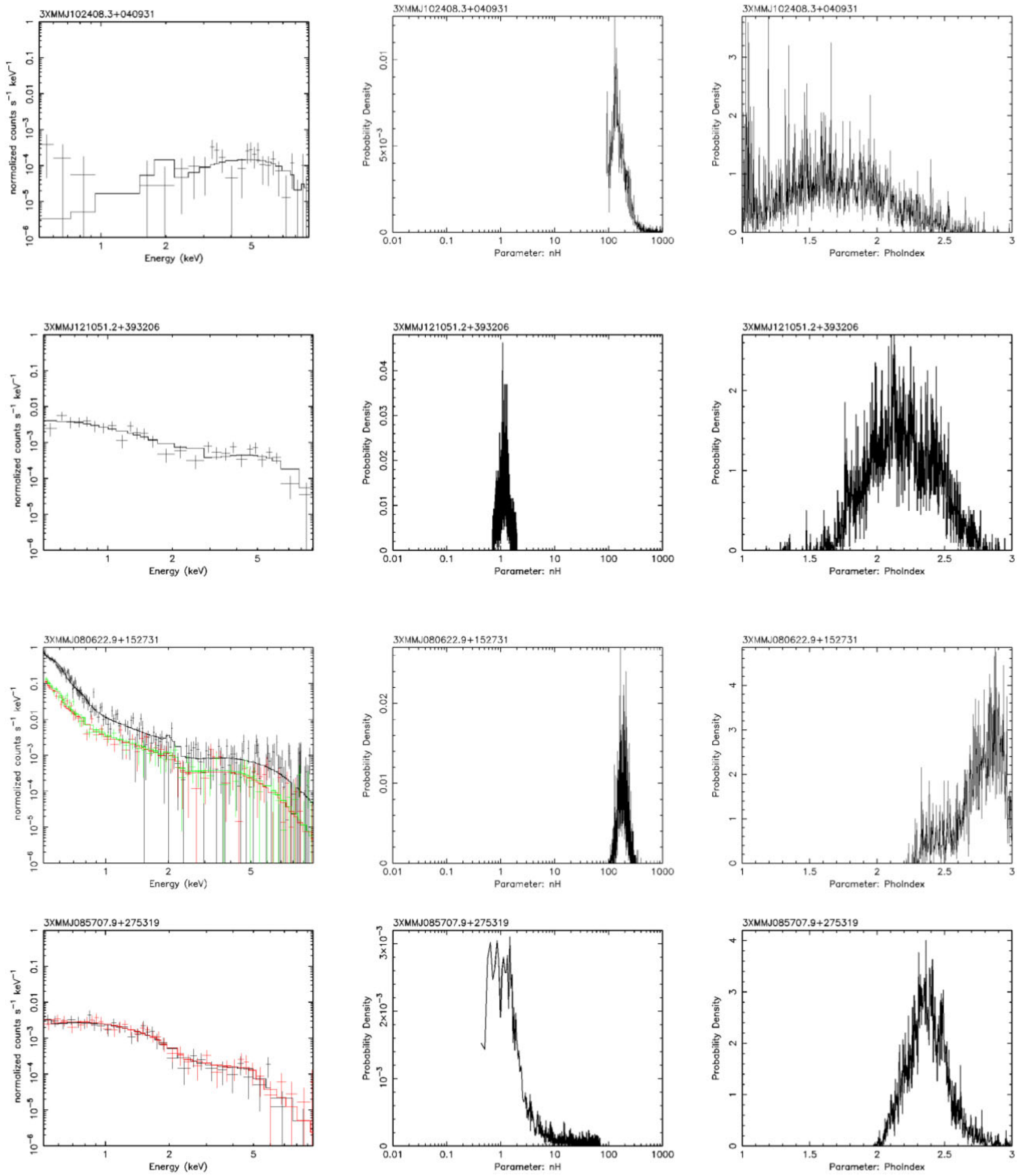

**Figure A1** The folded spectrum, probability distribution of column density ($\times 10^{24}$ cm$^{-2}$), and probability distribution of photon index, respectively, of 47 CT sources selected by MCMC method in our *bona fide* CT Torus-selected sample.

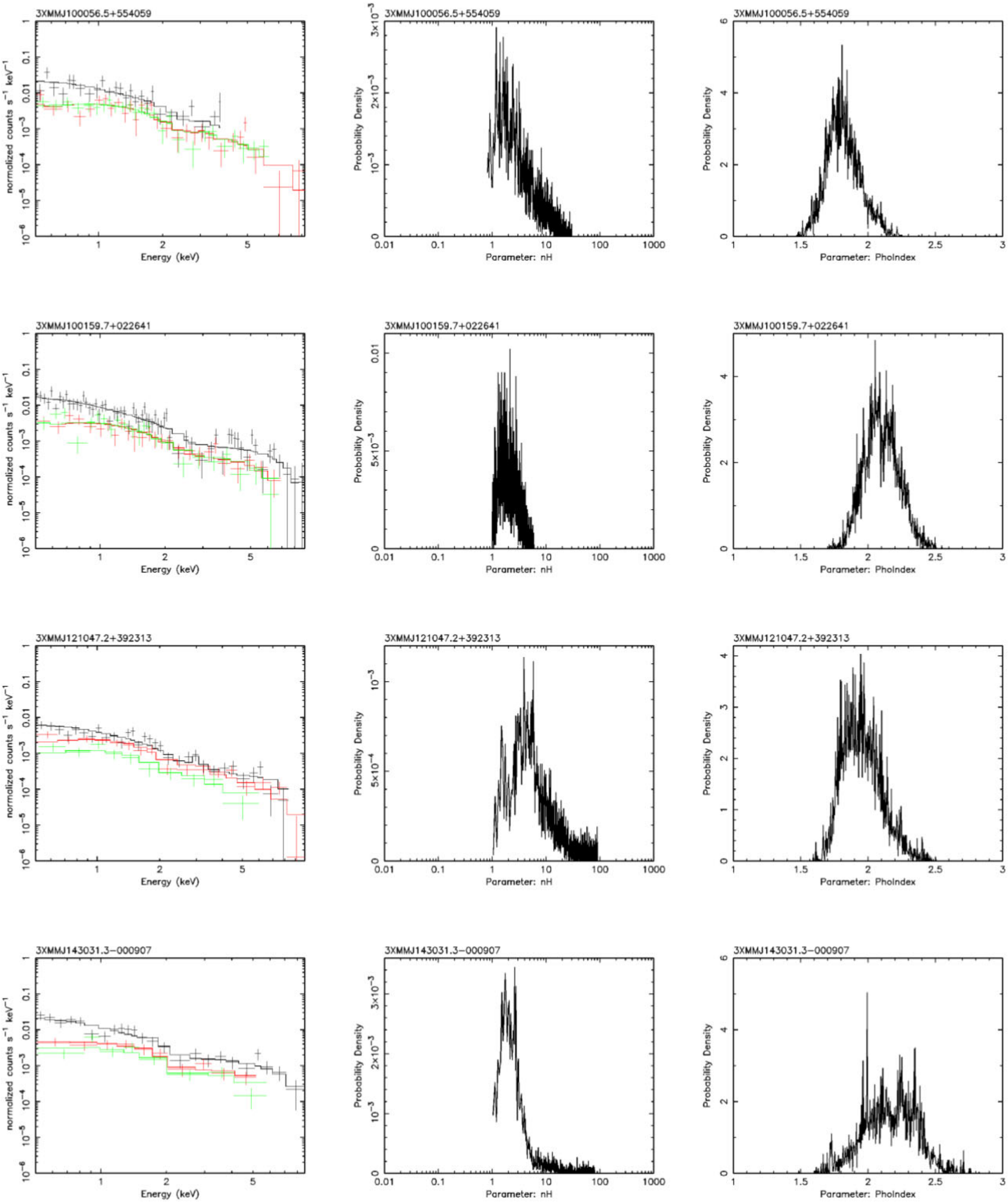

**Figure A1** – *continued*

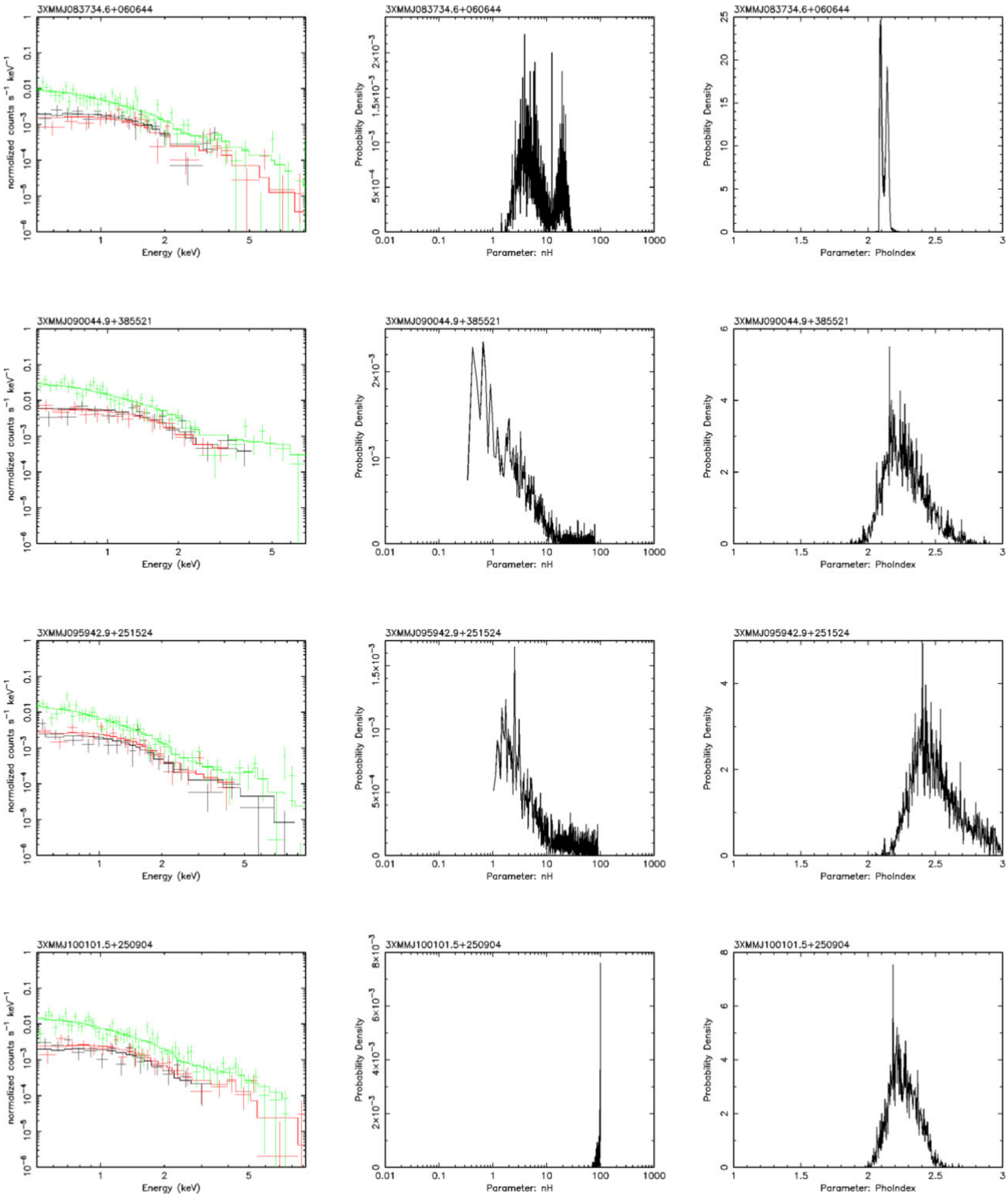

**Figure A1** – *continued*

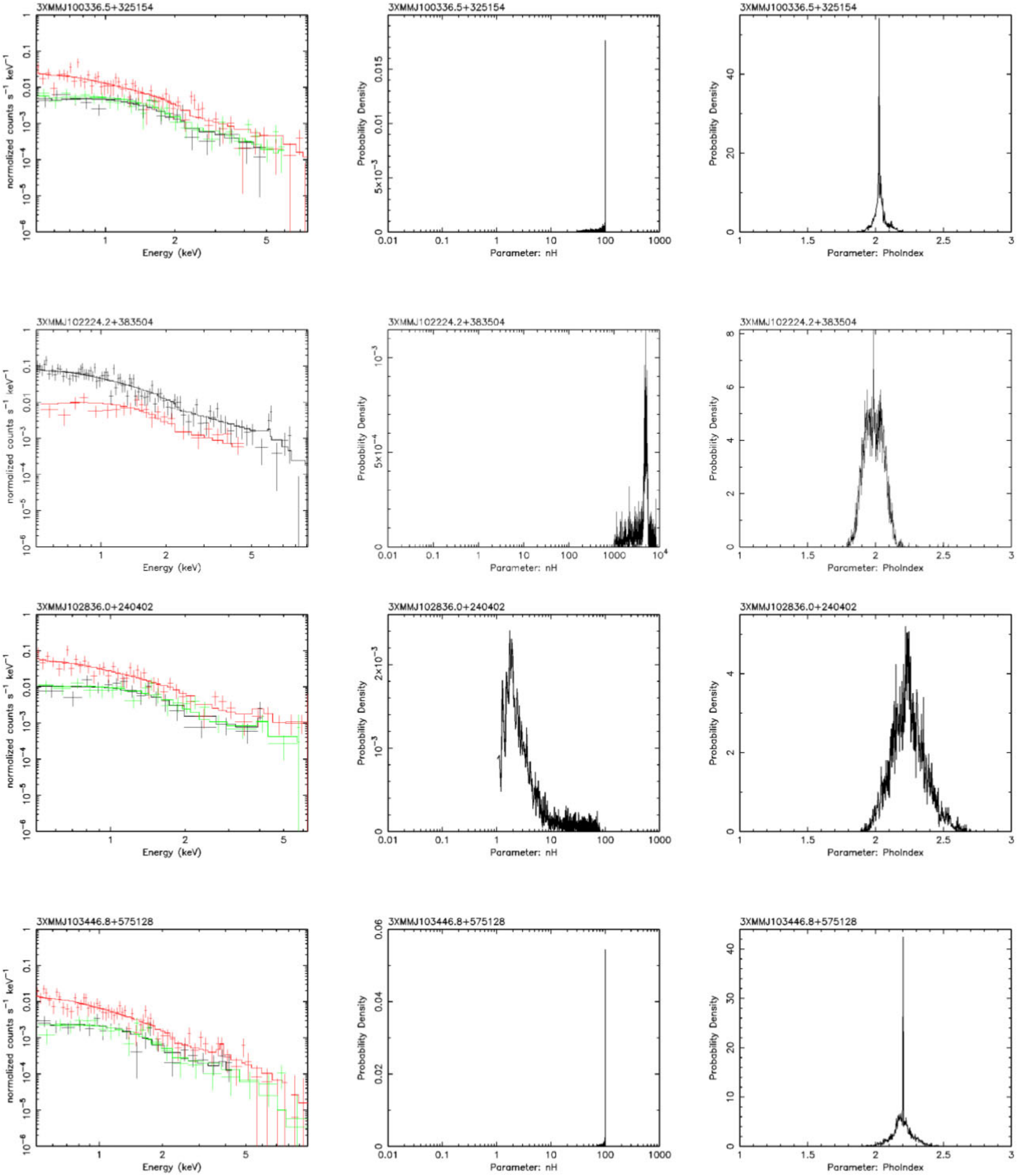

**Figure A1** – *continued*

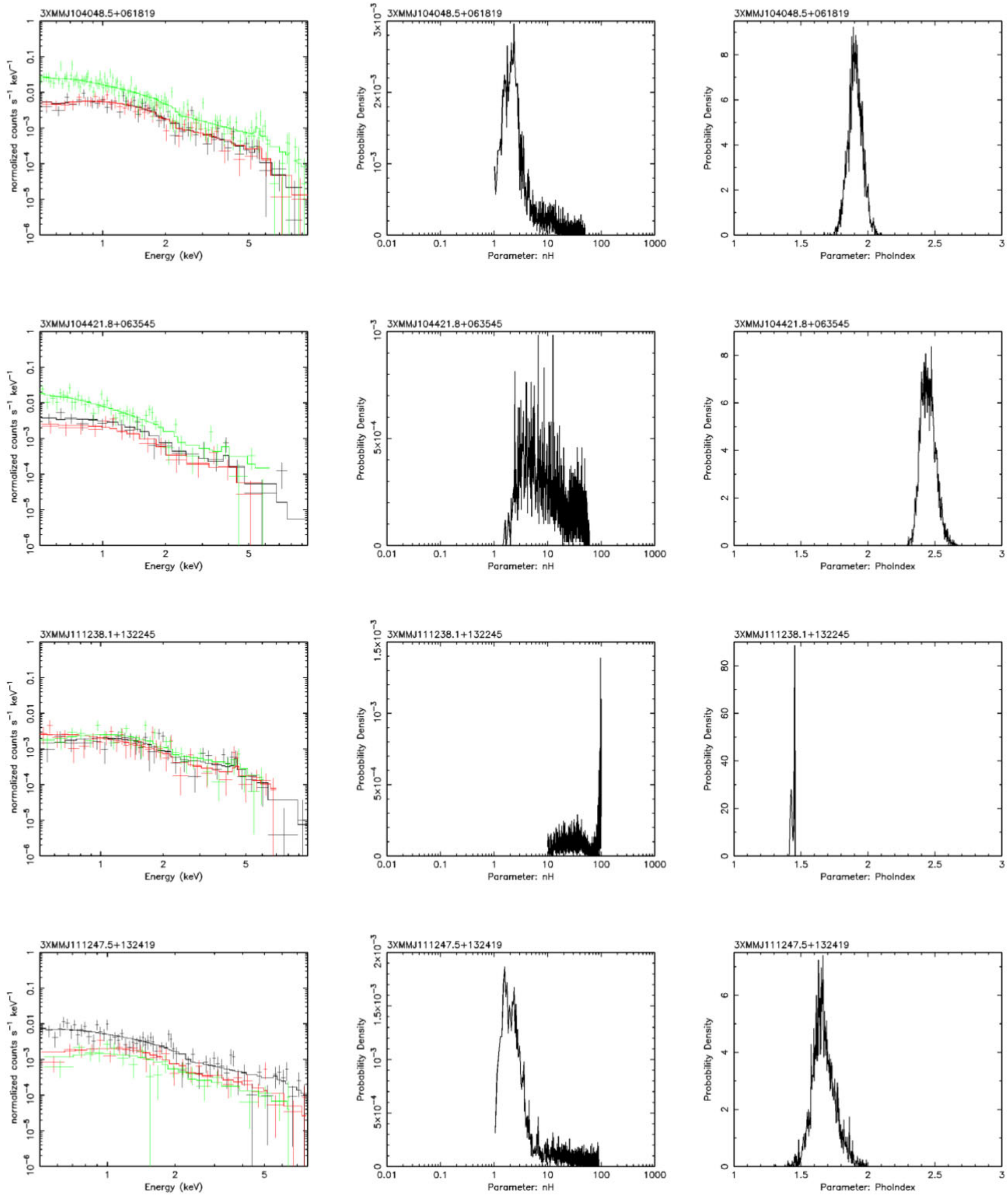

**Figure A1** – *continued*

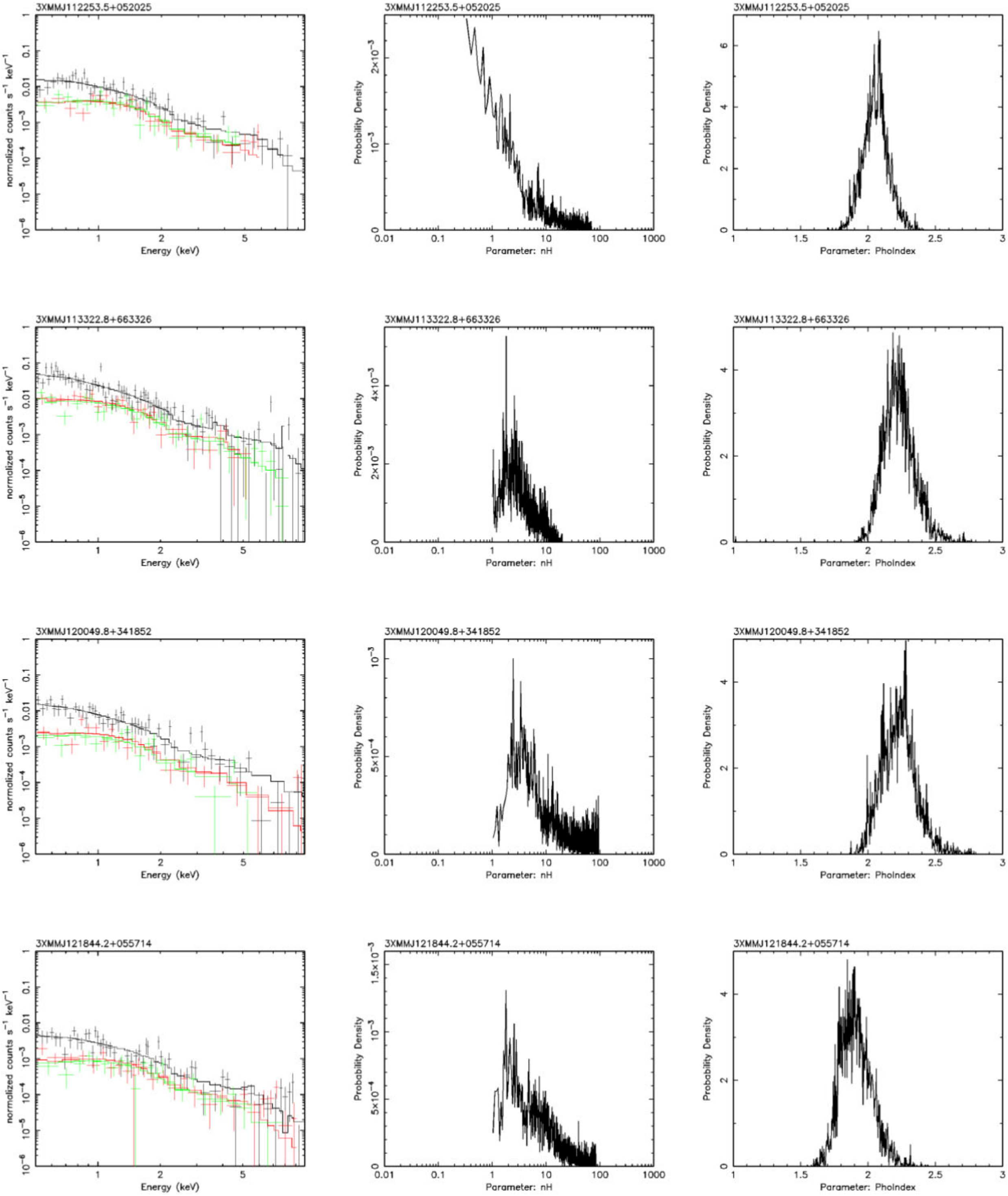

**Figure A1** – *continued*

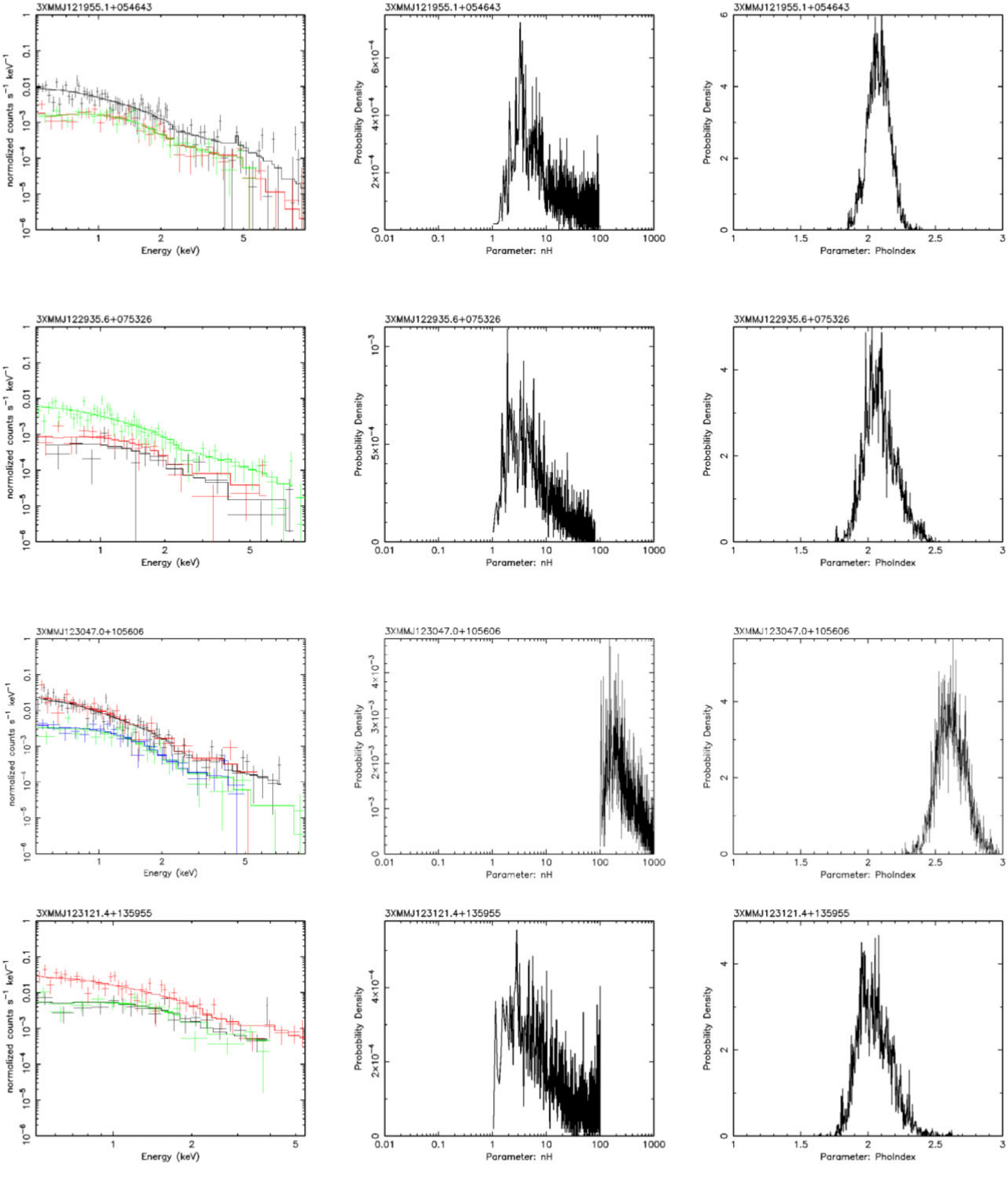

**Figure A1** – *continued*

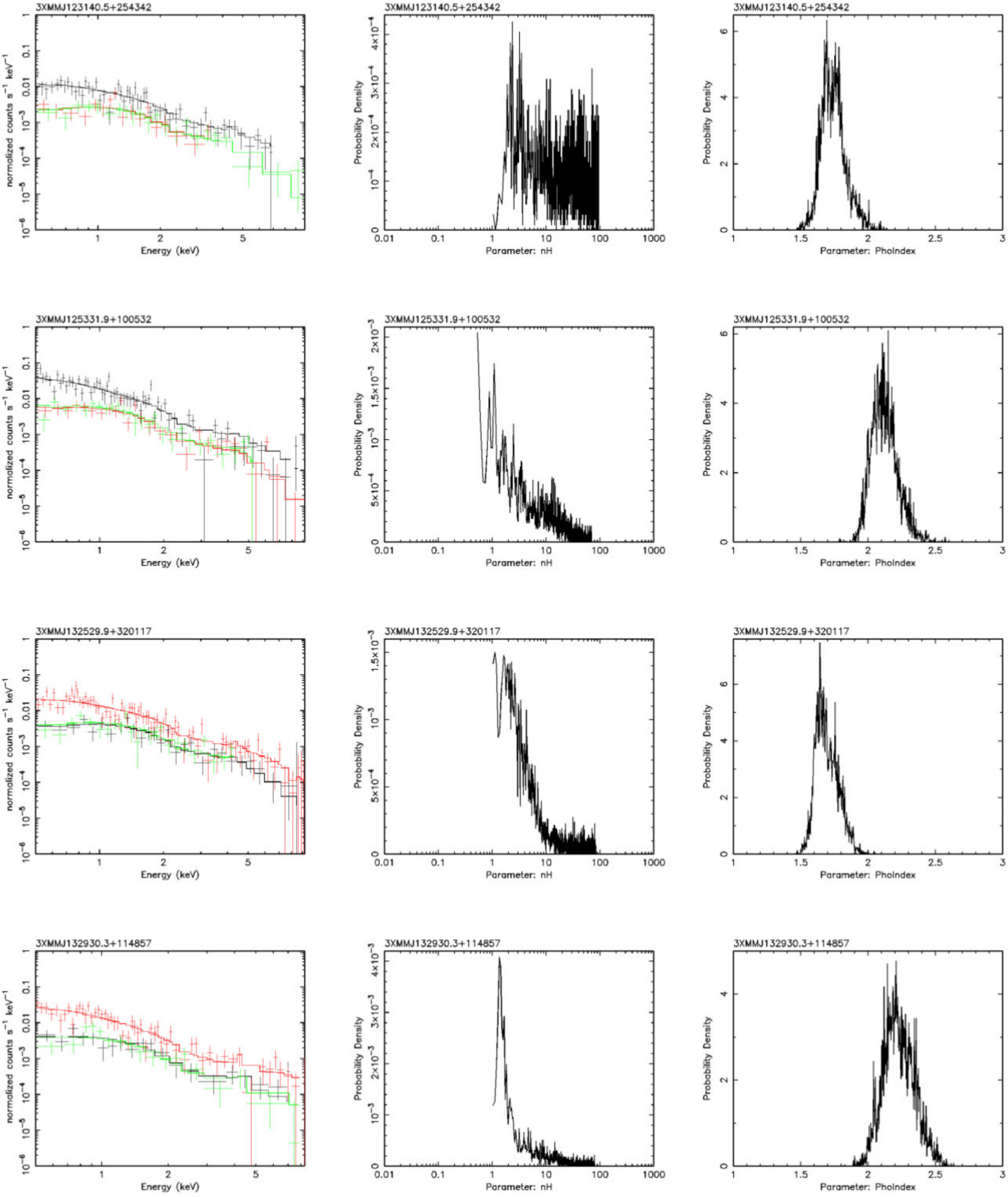

**Figure A1** – *continued*

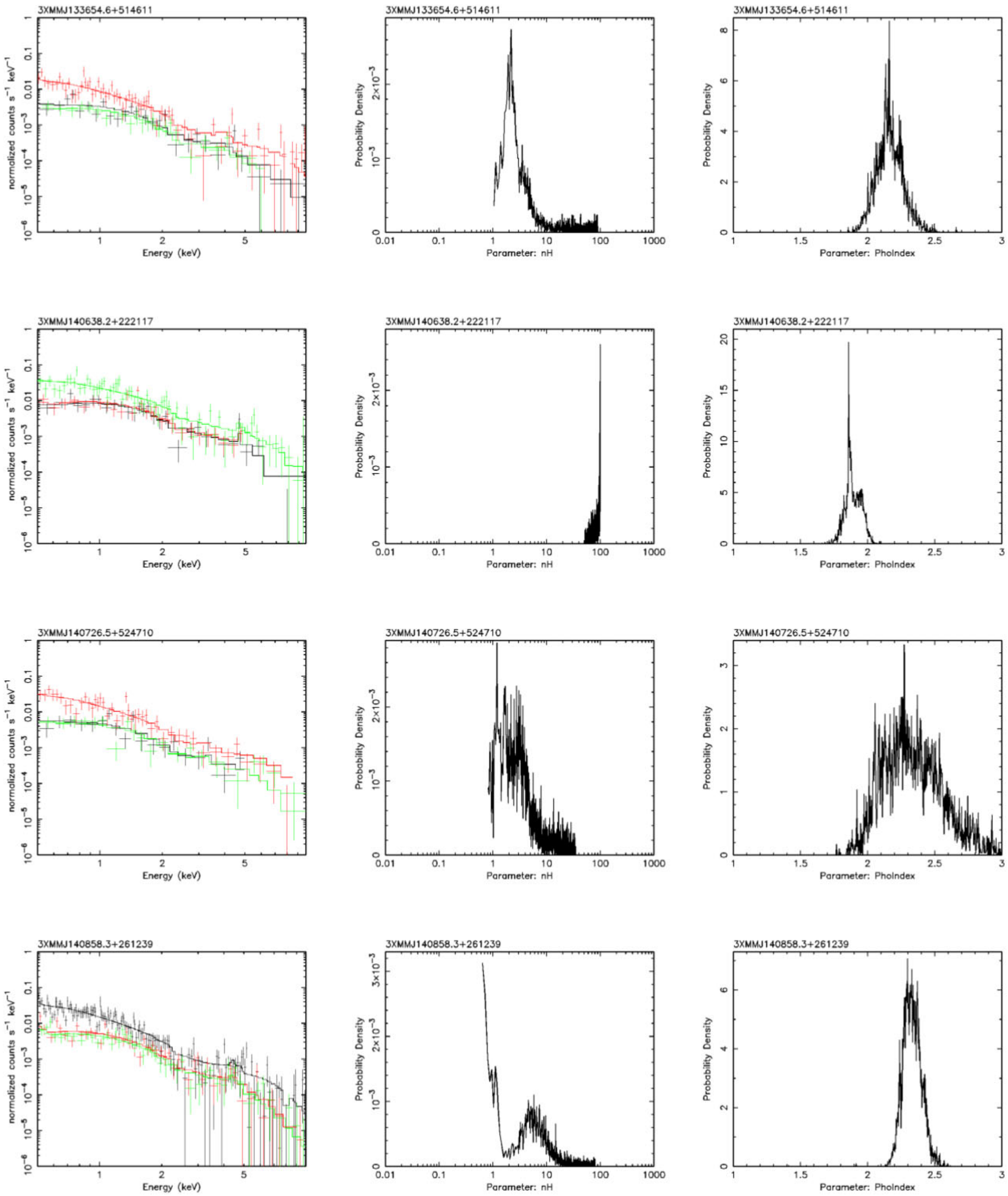

**Figure A1** – *continued*

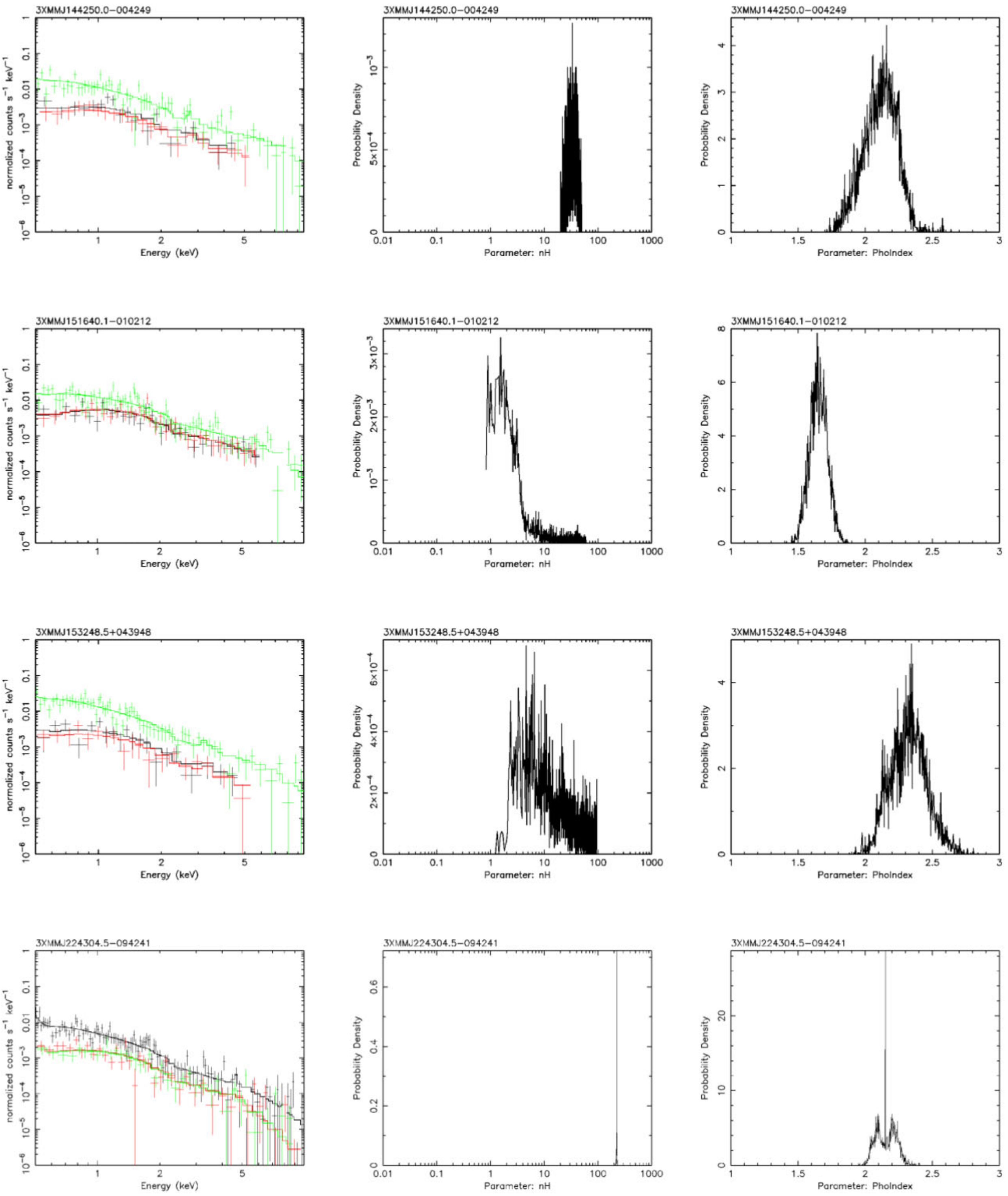

**Figure A1** – *continued*

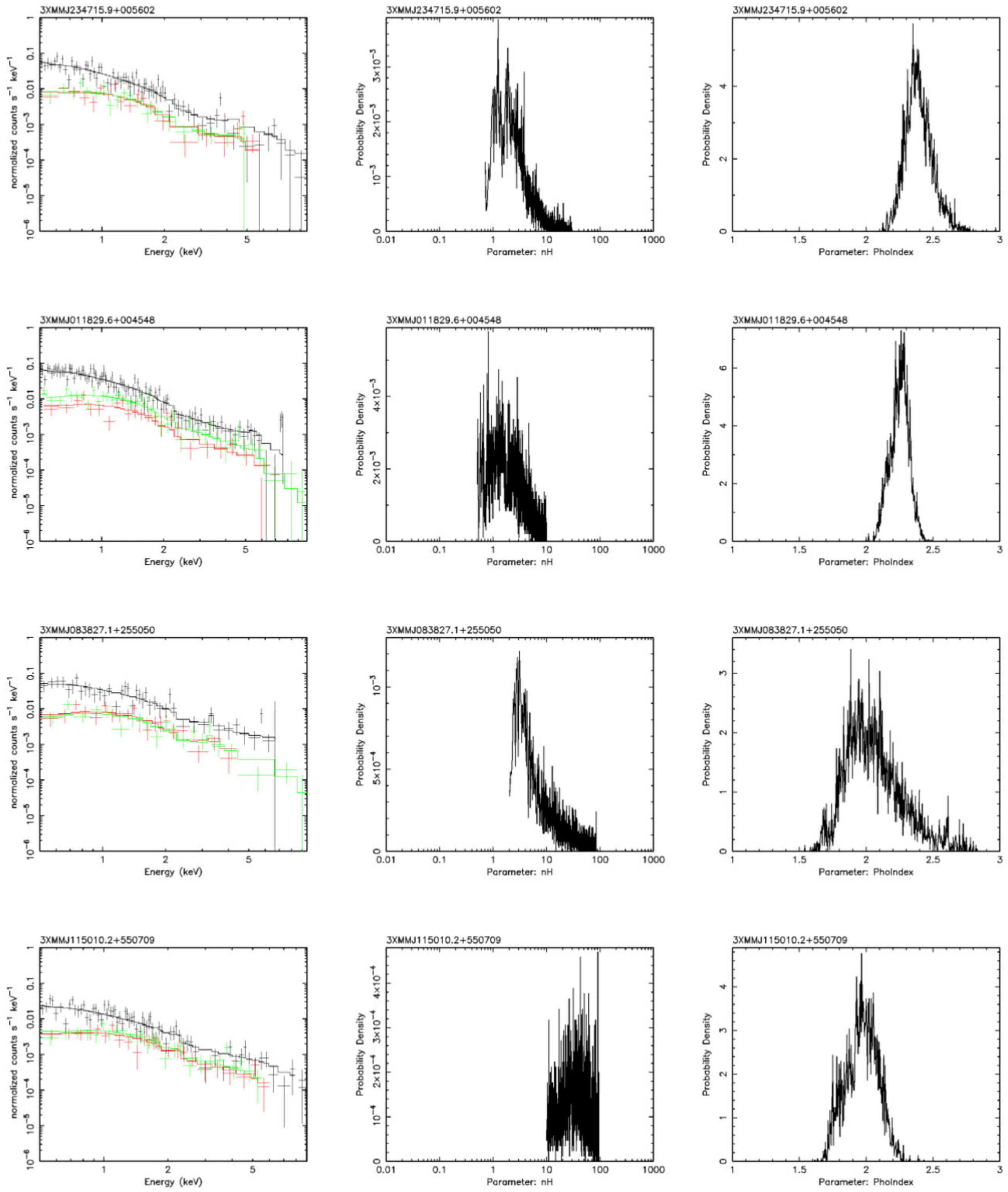

**Figure A1** – *continued*

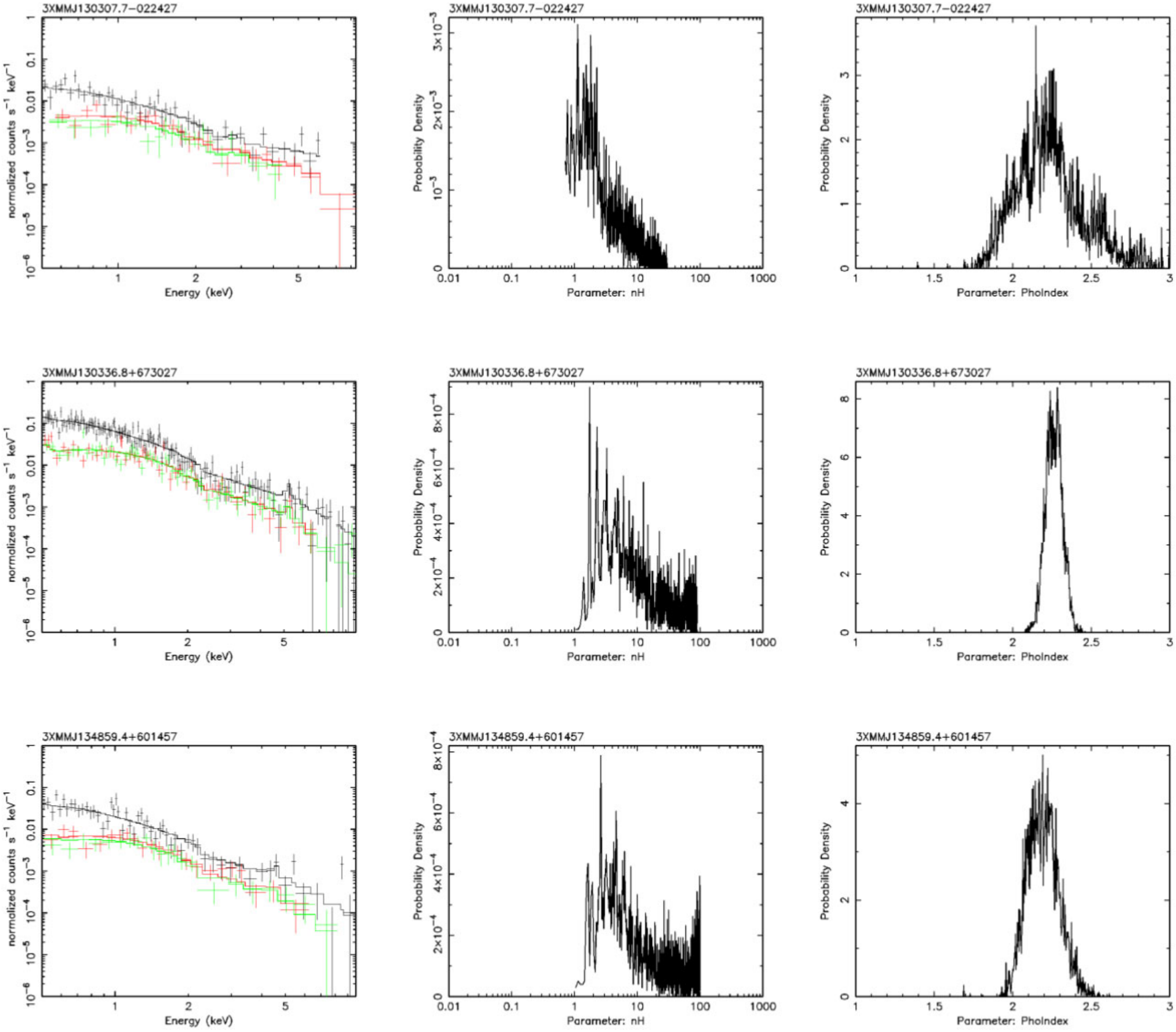

**Figure A1** – *continued*

This paper has been typeset from a TEX/LATEX file prepared by the author.